\documentclass[]{emulateapj}
\usepackage{epsfig, natbib, graphicx, color,bm}

\input epsf 

\usepackage{color}

\newcommand{\tablecomment}[1]{\let\thefootnote\relax\footnotetext{#1}}

\newcommand{\Mpc}{\mbox{Mpc}}

\newcommand{\msun}{M_\odot}

\newcommand{\nn}{\nonumber}

\newcommand{\fgas}{f_{gas}}

\newcommand{\avg}[1]{\left\langle #1 \right\rangle}
\newcommand{\Var}{\mbox{Var}}

\newcommand{\Ysz}{Y_{\rm SZ}}

\newcommand{\Lx}{L_{\rm X}}

\newcommand{\be}{\begin{equation}}
\newcommand{\ee}{\end{equation}}
\newcommand{\bea}{\begin{eqnarray}}
\newcommand{\eea}{\end{eqnarray}}

\newcommand{\keV}{\mbox{keV}}

\newcommand{\rosat}{{\it ROSAT}}
\newcommand{\chandra}{{\it Chandra}}
\newcommand{\xmm}{{\it XMM}}
\newcommand{\planck}{{\it Planck}}

\newcommand{\Rf}{R_{500}}

\newcommand{\Nt}{N_{200}}

\newcommand{\Atsz}{A_{tSZ}}

\newcommand{\Mwl}{M_{wl}}

\shortauthors{Rozo et al.}
\shorttitle{A Self-Consistent Model of Cluster Scaling Relations}

\begin{document}
\title{Closing the Loop: A Self-Consistent Model of Optical, X-ray, and SZ Scaling Relations for Clusters of Galaxies}
\author{Eduardo Rozo\altaffilmark{1,2}, James G. Bartlett\altaffilmark{3,4}, August E. Evrard\altaffilmark{5}, Eli S. Rykoff\altaffilmark{6,7}}
\altaffiltext{1}{Einstein Fellow, Department of Astronomy \& Astrophysics, The University of Chicago, Chicago, IL 60637.}
\altaffiltext{2}{Kavli Institute for Cosmological Physics, Chicago, IL 60637.}
\altaffiltext{3}{APC, AstroParticule et Cosmologie, Universit\'e Paris Diderot, CNRS/IN2P3, CEA/lrfu, Observatoire de Paris, Sorbonne Paris
Cit\'e, 10, rue Alice Domon et L\'eonie Duquet, Paris Cedex 13, France.}
\altaffiltext{4}{Jet Propulsion Laboratory, California Institute of Technology,
4800 Oak Grove Drive, Pasadena, CA, U.S.A.}
\altaffiltext{5}{Departments of Physics and Astronomy and Michigan Center for Theoretical Physics, University of Michigan, Ann Arbor, MI 48109.}
\altaffiltext{6}{SLAC National Accelerator Laboratory, Menlo Park, CA 94025.}
\altaffiltext{7}{Lawrence Berkeley National Laboratory, Berkeley, CA 94720.}

\begin{abstract}
We demonstrate that optical data from SDSS, X-ray data from \rosat\ and \chandra, and SZ data from
\planck, can be modeled in a fully self-consistent manner.  After accounting for systematic errors and allowing for property covariance, we find that scaling relations derived from optical and X-ray selected cluster samples  are consistent with one another.  Moreover,
these clusters scaling relations satisfy several non-trivial spatial abundance constraints and closure relations.   
Given the good agreement between optical and X-ray samples, we combine the two and derive
a joint set of $\Lx$--$M$ and $\Ysz$--$M$ relations.  Our best fit $\Ysz$--$M$ relation is in
good agreement with the observed amplitude of the thermal SZ power spectrum for a WMAP7 cosmology,
and is consistent with the masses for the two CLASH galaxy clusters published thus far.
We predict the halo masses of the remaining $z\leq 0.4$ CLASH clusters,
and use our scaling relations to compare our results with a variety of X-ray and weak lensing 
cluster masses from the literature.
\end{abstract}
 \keywords{
cosmology: clusters 
}

\section{Introduction}

Cluster scaling relations are of fundamental importance in both cosmological and astrophysical 
contexts.  From the cosmological point of view, galaxy clusters are potentially the most precise
probe of structure growth available today \citep[e.g.][]{cunhaetal09,oguritakada10,weinbergetal12}.
Given a cosmological model, robust theoretical predictions for the spatial abundance of massive halos as a function
of mass \citep[e.g.][]{jenkinsetal01,shethtormen02,warrenetal06,tinkeretal08,bhattacharyaetal11} are combined with cluster scaling relations to model the space-time abundance
of galaxy clusters as a function of their observable properties ({\sl e.g.}, X-ray luminosity, $\Lx$, 
Sunyaev-Zel'dovich (SZ) signal, $\Ysz$, optical richness, $\Nt$).  Such an approach has enabled recent cosmological 
constraints from cluster survey counts at a variety of 
wavelengths \citep[e.g.][]{henryetal09,vikhlininetal09,mantzetal10a,rozoetal10a,bensonetal11,sehgaletal11}.

From an astrophysical perspective, the self-similar model of \citet{kaiser86} makes concrete predictions
for how cluster scaling relations should behave in the absence of non-gravitational physical processes.  The specific predictions of this model were quickly shown to be a poor match to the local X-ray cluster luminosity function \citep{evrardHenry91, kaiser91}, but the underlying notion that massive halos are {\sl nearly} self-similar remains well supported by current data and numerical simulations \citep[][and references therein.]{allenetal11}.   

The intrinsic scaling behavior of massive halos across space-time is formally a theoretical construct, independent of observations.  
That is, given two quantities $\psi$ and $\chi$, the conditional probability distribution $P(\ln \psi|\chi,z)$ 
is independent of cluster selection {\it by definition}.  Of course, in practice, the distribution of clusters in the $\psi$--$\chi$
plane of any given cluster sample 
must necessarily reflect not only the intrinsic probability distribution $P(\ln \psi|\chi,z)$, but also how the clusters were
selected in the first place.   For instance, 
if a $\chi$-selected sample is complete above some threshold $\chi_{min}$, the for any $\chi$-bin with $\chi \geq \chi_{min}$,
the sample mean of $\ln \psi$ is an unbiased estimator of $\avg{\ln \psi|\chi}$.
By contrast, the sample mean of $\ln \chi$ for galaxy clusters in a given $\psi$-bin is not an unbiased estimator
of $\avg{\ln \chi|\psi}$; relative to a complete sample of $\psi$ selected clusters, the $\chi_{min}$ limit removes
systems with $\chi \leq \chi_{min}$.  Consequently, this selection effects must be explicitly accounted for when 
estimating $\avg{\ln \psi|\chi}$.
The job of observers is
to model these selection effects, relating the observed distribution back to selection-independent quantities like $P(\ln \psi|\chi,z)$.
In so doing, one can fairly compare the results from different cluster samples, regardless of how they are selected.\footnote{In
a cosmological context, one can forward model any selection effects into the observed distribution of galaxy clusters in
the $\psi$--$\chi$ space.  However, the nuisance parameters should be those characterizing $P(\ln \psi|\chi,z)$ if one
is to be able to compare results from the different cluster catalogs, and/or relate these nuisance parameters to different
astrophysical processes.}

Here, we will {\it define} the $\psi$--$\chi$ scaling relation to be the proability distribution $P(\ln \psi|\chi,z)$.  This choice of definition
implies that, in the absence of systematics, the scaling relations estimated from optical, X-ray, and SZ cluster samples must necessarily
be consistent with one another.  Any tension them must necessarily reflect
either systematic errors in the data itself, or in the theoretical models used to relate the intrinsic probability distribution $P(\ln \psi|\chi,z)$
to the observed cluster population. 
In this context, we aim to resolve the tension generated by the results of \citet{planck11_optical} --- henceforth referred to as P11-opt --- on the SZ signal of maxBCG galaxy clusters \citep{koesteretal07a}.  

P11-opt predicted the $\Ysz$--$\Nt$ scaling relation for the maxBCG systems
using an X-ray model from \citet{arnaudetal10} and the scaling relations from \citet{rozoetal09a}, henceforth 
referred to as R09.
They found that their predicted $\Ysz$--$\Nt$ relation was much higher than the observed signal.
We provide an explanation for the P11-opt results that involves
both types of errors noted above: unaccounted systematic errors in the optical data, the X-ray data, and the SZ data, and an incorrect model employed by P11-opt to derive $\Ysz$ expectations under $\Nt$ selection.

Here, we construct a self-consistent model of cluster scaling relations that can adequately
explain the P11-opt data.  Because the model used to predict the $\Ysz$--$\Nt$ scaling relation 
also involves X-ray data, we consider not just the $\Ysz$--$\Nt$ scaling relation, but all SZ, $\Lx$, and optical
scaling relations that have been directly measured.   
We explicitly test whether cluster scaling relations derived
from optical, X-ray, and SZ catalogs are consistent with each other, 
and we demand that our scaling relations be consistent
with cosmological expectations for the currently favored flat $\Lambda$CDM cosmological
model from WMAP7 \citep{komatsuetal11}.  We further demand that the spatial abundance functions in each of 
the observables be consistent with each other.  

This set of tests and self-consistency checks is {\it necessary} 
to resolve the problem posed by the P11-opt data.  For instance, consider a basic 
interpretation of the P11-opt result as indicating that the mass scale of either optical galaxy clusters or X-ray clusters (or both) is incorrect.   Any changes to the mass scales of these objects necessarily alters the predicted spatial abundances within a WMAP7 cosmology, so there is a very real risk that ``fixing'' the $\Ysz$--$\Nt$ relation this way could ``break'' the predicted cosmological counts.  Similarly, such a fix would affect other scalings, {\sl e.g.}, $\Lx$--$\Nt$, $\Lx$--$M$, etc., and the true multivariate model must be internally consistent.  Broadly speaking, we must be able to close all possible loops in this multi-dimensional parameter space.   
The fact that scaling relations and cluster counts are so interconnected offers the potential to obtain precise constraints on multivariate scalings and cosmology from joint sample analysis \citep{cunha09}.  

This is the third and final paper of a series that intends to explain the tension originally pointed out by P11-opt.  
In \citet[][hereafter paper I]{rozoetal12b}, we use pairs of clusters common to independent X-ray samples to measure  systematic uncertainties in derived X-ray properties and inferred halo masses.  In \citet[][paper II]{rozoetal12c}, we use the results of paper I to characterize the impact of these systematic errors on X-ray cluster scaling relations.   Combined, papers I and II develop the necessary foundations for our treatment
of the $\Ysz$--$\Nt$ scaling relation, the main aim of this paper.

The paper is organized as follows.  In \S\ref{sec:data}, we briefly summarize the various data sets we employ.
In section \ref{sec:yszn200} we consider the $\Ysz$--$\Nt$ relation, and discuss what is necessary to ``fix'' it.
The remainder of the paper is focused on checking whether our fix of the $\Ysz$--$\Nt$ relation is consistent with all
other data connected to it.  In \S\ref{sec:spatial abundance} we consider whether our modified scaling relations
and mass calibrations are consistent with cosmological expectations, and whether the optical and X-ray
cluster spatial abundances are consistent with one another.  Section \ref{sec:optical_xray_SZ_consistency}
compares the $\Lx$--$M$ and $\Ysz$--$M$ scaling relations derived from optical and X-ray selected cluster
catalogs.
In addition, we also compare the amplitude of the thermal SZ power
spectrum predicted from our best fit $\Ysz$--$M$ scaling relation to measurements from the
South Pole Telescope \citep{reichardtetal11}.  In \S\ref{sec:self-consistency} we 
test the self-consistency of our scaling relations,
testing whether we can combine the $\Ysz$--$\Nt$ and $\Lx$--$\Nt$ scaling relation to derive the observed
$\Ysz$--$\Lx$ scaling relation.  In \S\ref{sec:maxbcg_masses} we take a closer look at the maxBCG
mass calibration, derive our set of preferred scaling relations by combining maxBCG and V09 scaling relations,
and then use these to predict masses for each of the CLASH systems with $z\leq 0.4$.  We also compare
our predicted masses with a broad range of works from the literature, and predict the amplitude of the
thermal Sunyaev--Zeldovich power spectrum, and compare it with observations.  
Section \ref{sec:discussion} presents a summary and discussion of our results.

When considering cosmological predictions for cluster spatial abundances, we adopt the best fit WMAP7+BAO+$H_0$
flat $\Lambda$CDM cosmology from \citet{komatsuetal11}: $\sigma_8=0.816$, $H_0=70.2\ \mbox{km/s/Mpc}$,
$\Omega_m=0.274$, and $n_s=0.968$.  Cluster scaling relations were estimated using flat $\Lambda$CDM
models with $h=0.7$, though there is some variance in the value of $\Omega_m$ between various works, with
$\Omega_m\in [0.27,0.3]$ depending on the work.  
The impact of this level of variation in the recovered cluster scaling relations is $\lesssim 5\%$.   The measurement conventions we employ are given in \S\ref{sec:conventions}


\section{Data Sets and Conventions}
\label{sec:data}

The data employed in this work is collated from a variety of papers cited in Papers I and II, and we direct the reader to these papers for details.  We provide a brief summary in this section.   

A note regarding mass is in order.  Throughout, we take halo mass, $M$, to be $M_{500c}$, the mass defined within an overdensity of 500 with respect to the critical density at the redshift of the cluster.   While the true mass is not strictly observable, it can be inferred from observable properties in a model-dependent manner.  The published X-ray samples we employ calibrate scaling relations using a subset of clusters with hydrostatic mass measurements, then derive mass estimates for the entire sample based on the implied scaling relation.   Thus, our use of mass, $M$, should generally be interpreted as an observable rescaled to provide an estimate of $M_{500c}$ (see Paper I).  Statistical inferences are unaffected by this complication, as the $M_{500c}$ estimates so derived are, by definition, statistically consistent with pure hydrostatic masses.

\subsection{Data Sets}

Our work with optically selected galaxy clusters is based on the Sloan Digital Sky Survey (SDSS) maxBCG cluster
catalog \citep{koesteretal07a}.  The optical observable is the red-sequence count of galaxies, also known as optical richness, $\Nt$.  
The $\Lx$--$\Nt$ and $M$--$\Nt$ scaling relations are those from 
\citet[][R09]{rozoetal09a}, who applied various corrections to the original set of scaling relations reported 
in \citet{rykoffetal08a} and \citet{johnstonetal07}.  The $\Ysz$--$\Nt$ scaling relation is that from the
P11-opt \citep{planck11_optical}, corrected for the effects of cluster miscentering, as
as per the results of \citet{biesiadzinskietal12}, and for the expected level of aperture-induced
measurement bias in the data (see \S\ref{sec:yszn200}).  These corrections are applied on the binned $\Ysz$
data and scaling relations quoted in P11-opt, and do not employ the raw \planck\ data in any way.

Turning now to X-ray selected systems, most of our work is based on the data presented in
\citet{vikhlininetal09}, henceforth referred to simply as V09.  The V09 galaxy clusters are X-ray selected,
and the cluster masses are estimated based on the $M$--$Y_X$ scaling relation, which is itself
calibrated using hydrostatic mass estimates as described in \citet{vikhlininetal06}.  These cluster
masses are used to estimate the $\Lx$--$M$ scaling relation, while the $\Ysz$--$M$ scaling relations is
derived from the $M$--$Y_X$ relation above with the $\Ysz$--$Y_X$ for the V09 clusters as quoted in
\citet{rozoetal12a}.  The derivation of the $\Ysz$--$M$ and $\Ysz$--$\Lx$ scaling relation for the
V09 data set is presented in paper II.

In section \ref{sec:yszn200}, we also consider the X-ray data from \citet{mantzetal10b}, as well as that
of \citet{planck11_local} (henceforth P11), \citet{arnaudetal10}, and \citet{prattetal09}.  In all cases,
mass calibration is obtained using hydrostatic mass 
estimates \citep[e.g.][]{pointecouteauetal05,arnaudetal07,allenetal08}.  Concerning the P11 data,
P11 distinguish between A and B systems on the basis of 
the apparent angular size of the cluster in the sky.  All
of the galaxy clusters in the redshift range of interest ($z\approx 0.2$ to match the maxBCG catalog) are
A clusters.  In paper I, we noted the X-ray data of clusters A and B are systematically different.
Since we are interested in the comparison to maxBCG galaxy clusters, we follow our work on 
papers I and II, and focus exclusively on the  P11 data
for systems in the redshift slice $z\in[0.12,0.3]$, and use P11(z=0.23) to denote this cluster subsample. 
We refer the reader to papers I and II for further details.

Turning to the $\Ysz$--$\Lx$ scaling relation, we consider data derived both from the P11(z=0.23) data set,
as well as the measurement reported in \citet{planck11_xray}, which we denote P11-X.  
P11-X estimates the $\Ysz$--$\Lx$ scaling
relation by stacking X-ray selected clusters from the compilation MCXC catalog \citep{piffarettietal11}.


\subsection{Conventions}\label{sec:conventions}

The optical richness, $\Nt$, measures the number of galaxies within a $g-r$ color-cut around
the color of the galaxy designated as the brightest cluster galaxy, 
and within a scaled radial aperture.\footnote{Despite the subscript, this aperture is not a good estimator for the
$R_{200}$ of the cluster, and should only be considered as an optical ``tag'' that correlated with mass.}
X-ray luminosity, $\Lx$, is defined as the rest-frame $[0.1,2.4]\ \keV$ band luminosity within a cylindrical aperture of $\Rf$.  V09 has a somewhat different definition, as does R09, but we correct for these differences (Paper I).  We quote luminosity in units of $10^{44}\ \mbox{ergs/s}$.
Finally, $\Ysz$ is defined as the integrated SZ signal within the cluster radius $\Rf$.  
In some cases, such as the analysis in P11-opt, the aperture $\Rf$ used to estimate $\Ysz$
explicitly depends on the richness $\Nt$, which results in modest aperture-induced
corrections.  This is discussed in more detail in Section \ref{sec:yszn200}, and is also
relevant to the discussion in section \ref{sec:self-consistency}.   Because the
scaling relation $\Ysz$--$\Nt$ explicitly depends on the angular diameter distance, we always
quote scaling relations with respect to $D_A^2\Ysz$ rather than $\Ysz$ alone.  For instance,
we quote the amplitudes for the scaling relations $D_A^2\Ysz$--$N$ and $D_A^2\Ysz$--$\Lx$ rather 
than $\Ysz$--$\Nt$ or $\Ysz$--$\Lx$.  However, for brevity, we will refer to these $D_A^2$ weighted
relations as the $\Ysz$--$\Nt$ and $\Ysz$--$\Lx$ scaling relations: the $D_A^2$ factor will often
be implied.  In all cases, we measure $D_A^2\Ysz$ in units of the $10^{-5}\ \Mpc^2$.

Turning to scaling relations, given two arbitrary cluster observables $\psi$ and $\chi$, we assume the
probability distribution $P(\psi|\chi)$ is a log-normal distribution.   As in paper II, we model the mean
of this distribution as a linear relation in log-space, so that
\be
\avg{\ln \psi|\chi} = a_{\psi|\chi} + \alpha_{\psi|\chi} \ln (\chi/\chi_0) \label{eq:mean}
\ee
where $a_{\psi|\chi}$ is the amplitude parameter, and $\alpha_{\psi|\chi}$ is the slope.  The parameter $\chi_0$
is the pivot point of the relation, which we always select so as to decorrelate the amplitude and slope
parameters.  The variance in $\ln \psi$ at fixed $\chi$ is assumed to be constant, and is denoted as
\be
\Var(\ln \psi|\chi) \equiv \sigma_{\psi|\chi}^2.
\ee

We note that binned data naturally measures the moment $\avg{\psi|\chi}$ rather
than $\avg{\ln \psi|\chi}$.  The two are related via
\be
\ln \left( \avg{\psi|\chi} \right) = \avg{\ln \psi|\chi} + \frac{1}{2}\sigma_{\psi|\chi}^2.
\ee
We define $\tilde a_{\psi|\chi}$ as the value of $\ln \avg{\psi|\chi}$ at the pivot of the scaling relation, so
that
\be
\tilde a_{\psi|\chi} = a_{\psi|\chi} +  \frac{1}{2}\sigma_{\psi|\chi}^2.
\ee

We note that  if the scatter, $\sigma_{\psi|\chi}$, is correlated with either the amplitude, $a_{\psi|\chi}$, or slope,  $\alpha_{\psi|\chi}$, of the scaling relation, as it often is, then a pivot point that decorrelates $a_{\psi|\chi}$ and $\alpha_{\psi|\chi}$ need not decorrelate the mean linear amplitude, $\tilde a_{\psi|\chi}$, and slope, $\alpha_{\psi|\chi}$.  We explicitly take into account the difference between $a_{\psi|\chi}$ and $\tilde a_{\psi|\chi}$ in all of the comparisons performed in this paper.

To avoid any possible confusion, the subscripts we employ in this work are as follows: $m$ for mass, 
$n$ for $\Nt$, $x$ for $\Lx$, and $sz$ for $D_A^2\Ysz$.  So, for instance, $a_{sz|m}$ is the amplitude
of the $D_A^2\Ysz$--$M$ relation.


\section{The $\Ysz$--$N_{200}$ Scaling Relation}
\label{sec:yszn200}

\subsection{Data and Model Predictions}

We begin our investigation by exploring the $\Ysz$--$\Nt$ scaling relation measured by P11-opt. 
The data is taken directly from that work, and is corrected for the effects of cluster miscentering
as per the results of \citet{biesiadzinskietal12} \citep[see also][]{anguloetal12}.  The centering correction
applied is summarized in Table \ref{tab:yszn_data}, along with the corrected data points.  
We have also added in quadrature the uncertainty associated with the miscentering corrections 
to the error budget of P11-opt.  
We note that the errors in P11-opt are in fact asymmetrical, but we have opted to symmetrize them to simplify our analysis.
Finally, we also reduce the observed amplitude by $0.05$ to account for the Malmquist bias caused by aperture-induced 
covariance in the $\Ysz$ measurements (see below). 


\begin{deluxetable}{lll}
\tablewidth{0pt}
\tablecaption{$\Ysz$--$\Nt$ Data}
\tablecomment{
The data in the first two columns is from \citet[][P11-opt]{planck11_optical}, after being corrected for the effects of cluster 
miscentering following \citet{biesiadzinskietal12} (third column).  The uncertainty in the corrections is 
added in quadrature to the observational errors.  
\vspace{0.1in}
}
\tablehead{
$\Nt$ & $D_A^2\Ysz/(10^{-5}\ \Mpc^{-2})$ & Centering Correction}
\startdata
      10--      13 &     0.058$\pm$    0.012 &      0.74$\pm$     0.13 \\
      14--      17 &      0.107$\pm$    0.020 &      0.77$\pm$     0.11 \\
      18--      24 &      0.222$\pm$    0.028 &      0.80$\pm$    0.08 \\
      25--      32 &      0.394$\pm$    0.044 &      0.82$\pm$    0.08 \\
      33--      43 &      0.692$\pm$    0.074 &      0.84$\pm$    0.07 \\
      44--      58 &       1.205$\pm$     0.130 &      0.86$\pm$    0.07 \\
      59--      77 &       1.876$\pm$     0.241 &      0.87$\pm$    0.07 \\
      78--     104 &       4.594$\pm$      1.009 &      0.89$\pm$    0.09 
\enddata
\label{tab:yszn_data}
\vspace{0.02in}
\end{deluxetable}



\begin{deluxetable*}{lllclll}
\tablewidth{0pt}
\tablecaption{Input ($\Ysz$--$M$ and $M$--$\Nt$) and Derived ($\Ysz$--$\Nt$) Scaling Relations at $z = 0.23$}
\tablecomment{
We assume the form $\avg{\ln \psi|\chi}= \ln \psi_0+\alpha \ln(\chi/\chi_0)$.  Units are $10^{14}\ \msun$ for mass, and $10^{-5}\ \Mpc^2$ for $D_A^2\Ysz$.   The quantity $\beta$ is the slope
of the halo mass function at the pivot scale of the $\Ysz$--$M$ relation.  
The $\Ysz$--$\Nt$ relations are derived by combining each $\Ysz$--$M$
relation with the R09 $M$--$\Nt$ relation.  
}
\tablehead{
Relation & $\chi_0$ & $\beta$ & $a_{sz|m}$, $\tilde a_{m|n}$, or $a_{sz|n}$ & $\alpha$ & $\sigma_{\ln \psi|\chi}$ & Sample}
\startdata
$D_A^2\Ysz$--$M$ & 4.8 & $2.75$ & $1.34\pm 0.07$$^a$ & $1.61\pm 0.11$ & $0.12\pm 0.03$ & V09 \\
$D_A^2\Ysz$--$M$ & 5.5 & $2.93$ &  $1.97\pm 0.06$ & $1.48\pm 0.12$ & $0.20\pm 0.04$ & P11(z=0.23) \\
$D_A^2\Ysz$--$M$ & 10.0 & $3.95$ &  $2.54\pm0.20$ & $1.48 \pm 0.09$ & $0.15\pm 0.03$ & M10  
\vspace{0.05in} \\
\hline \vspace{-0.05in} \\
$M$--$\Nt$ & 40 & --- & $0.95\pm 0.07$\footnote{The systematic error in the amplitudes ($\sim \pm 0.1$) are considered independently.} 
	& $1.06\pm 0.08\ (stat)\ \pm 0.08\ (sys)$ & $0.45\pm 0.1$ & maxBCG  
\vspace{0.05in} \\
\hline 
\hline \vspace{-0.05in} \\
$D_A^2\Ysz$--$\Nt$\footnote{We emphasize that even though we plot $\ln \avg{D_A^2\Ysz|\Nt}$ so as to match the P11-opt
data, the amplitude reported here is $a_{sz|n}$ rather than $\tilde a_{sz|n}$ (see text).}
	 & 70.0 & --- &  $1.13\pm 0.17$ & $1.70\pm 0.18$ & $0.78\pm 0.19$ & V09 \\
$D_A^2\Ysz$--$\Nt$$^b$ & 70.0 & --- &  $1.57\pm 0.16$ & $1.57\pm 0.18$ & $0.74\pm 0.18$ & P11(z=0.23) \\
$D_A^2\Ysz$--$\Nt$$^b$ & 70.0 & --- &  $1.26\pm 0.26$ & $1.57\pm 0.16$ & $0.73\pm 0.17$  & M10  \vspace{0.05in}
\enddata
\label{tab:rel1}
\end{deluxetable*}


We compare the P11-opt data to the predicted $\Ysz$--$\Nt$ relation obtained from combining
the $M$--$\Nt$ scaling relation from R09, and the
$\Ysz$--$M$ relations for the V09, M10, and P11(z=0.23) data sets derived in paper II.
These input scaling
relations are summarized in Table \ref{tab:rel1}, along with the predicted $\Ysz$--$\Nt$ relations.
We note that R09 report not the amplitude parameter $a_{m|n}$, but rather the parameter $\tilde a_{m|n}$
that characterizes the mean $\avg{M|\Nt}$.  
A similar caveat holds for $\Ysz$--$\Nt$, as \citet{planck11_optical}  compute $\avg{\Ysz|\Nt}$ rather than $\avg{\ln \Ysz|\Nt}$.
In this section, we will only consider the statistical uncertainty in the 
amplitude of the $M$--$\Nt$ relation;  the systematic error will be considered independently
below.

We use the formalism from Paper II to predict the $\Ysz$--$N_{200}$ scaling relation from these input scaling relations.   The amplitude and slope are given by
\bea
a_{sz|n} & = & \left[ a_{sz|m} + \alpha_{sz|m}a_{m|n} \right]  + r \beta \alpha_{sz|m}\sigma_{m|sz}\sigma_{m|n}  \label{eq:yszn_amp} , \\
\alpha_{sz|n} & = & \alpha_{sz|m}\alpha_{m|n} ,
\eea
where $r \equiv  r_{sz,n|m}$ is the correlation coefficient of $\Ysz$ and $\Nt$ at fixed halo mass,
and $-\beta$ is the local logarithmic slope of the cosmic mass function.  
The scatter in the mass--richness relation $\sigma_{m|n}$ is large ($0.45\pm 0.1$), primarily 
reflecting a poor choice
of richness estimator rather than a large intrinsic cluster variance \citep{rozoetal09b,rykoffetal12}.  Consequently, we do not expect the richness and SZ scatter to be strongly correlated at fixed mass.
For $r_{sz,n|m}=0.1$, $\beta=3$, $\sigma_{m|sz}=0.45$
and $\sigma_{m|sz}=0.1$, we find that the amplitude correlation term above is $\approx 2\%$, much too small
to be of relevance for this study.  

On the other hand, 
in a recent work, \citet{anguloetal12} used the Millenium-XXL simulation to estimate
the correlation coefficient between $\Ysz$ and $\Nt$, finding a value of
$r=0.47$.   We expect the large correlation coefficient of \citet{anguloetal12} reflects aperture effects. Specifically,
we wish to predict $\Ysz$--$\Nt$ when $\Ysz$ is measured within $\Rf$, whereas \citet{anguloetal12} explicitly measure
$\Ysz$ within the radius $\avg{\Rf|\Nt}$.
We can estimate the correlation coefficient quoted in \citet{anguloetal12} by noting that the scatter of the $\Ysz$--$M$ scaling
relation from V09 is $\sigma_{sz|m}=0.12$.  We also know from paper I that a bias 
$b_m$ in the observed mass induces
a bias of $\sim \! 0.3\ln b_m$ in $\ln \Ysz$ due to aperture effects.  Since the optical scatter in mass is $\sigma_{m|n}\approx 0.4$,
the induced scatter in $\Ysz$ via aperture effects is $\avg{\delta_{sz,induced}^2}^{1/2} \approx 0.3\times 0.4 \approx 0.1$.   This scatter is perfectly correlated with richness, while the total scatter is the intrinsic scatter plus the induced scatter added in quadrature.  The resulting, aperture-induced correlation coefficient is 
\bea
r & = & \frac{\avg{(\delta_{sz,int}+\delta_{sz,induced})\delta_n}}{\sigma_{sz|m}\sigma_{n|m}} \nn \\
	& \approx & \frac{0.1\times 0.4}{(0.1^2+0.12^2)^{1/2}\times 0.4} \approx 0.6,
\eea
in reasonable agreement with $r=0.47$ quoted in \citet{anguloetal12}.  
The total scatter (intrinsic + induced) is $\approx 0.16$, also in reasonable 
agreement with \citet{anguloetal12} who find $\sigma_{sz|m}=0.18$.

So which correlation coefficient should one adopt? If one wishes to predict $\Ysz$--$\Nt$ where $\Ysz$ is measured within $\Rf$,
as we do, then one should set $r_{sz,n|m}=0$, since $\Ysz$ within $\Rf$ is not subject to aperture effects.
That said, when we compare our predictions to the P11-opt data, we need to correct the P11-opt data for the
aperture-induced effect estimated above.  Thus, even though we set
$r_{sz,n|m}=0$ in our analysis, our comparison to data explicitly takes into account the impact of aperture-induced covariance.
We note, however, that
the fact that the P11-opt measurements are template-amplitude fits rather than cylindrically integrated $\Ysz$ measurements
will reduce the impact of said biases, as inner radii acquire higher statistical weight than in the case of a cylindrical
integration.  The naive bias estimate of $r\beta \alpha_{sz|m} \sigma_{m|sz}\sigma_{m|n}$
is $\approx 0.1$ for $r=0.5$, which sets an upper limit to the correction appropriate in the case of template amplitude fits.  
For our purposes, when comparing the P11-opt data to our predictions, we will decrease the P11-opt data by half of this amount,
and note that this correction has a $\pm 5\%$ systematic uncertainty in the amplitude of the $\Ysz$--$M$ 
relation.\footnote{There are some additional secondary effects that should reduce the correlation coefficient between
$\Ysz$ and $\Nt$ relative to the \citet{anguloetal12} measurement.  In particular, \citet{anguloetal12} base their predictions
on projecting the galaxy density field across the full simulation box, which is $4.11\ \mbox{Gpc}$.  This is to be compared to the comoving
width of the red-sequence, which is $\approx 0.1\ \mbox{Gpc}$.  In addition, the \citet{anguloetal12} measurement includes miscentering
induced covariance, whereas we are explicitly correcting the data for miscentering.  To the extent that the dominant effect
are the aperture-induced corrections, however, these differences should play a minor role.}

The scatter of the $\Ysz$--$\Nt$ relation is given by
\be
\sigma_{sz|n}^2 = \alpha_{sz|m}^2 \left[ \sigma_{m|n}^2 + \sigma_{m|sz}^2 - 2r_{sz,n|m}\sigma_{m|n}\sigma_{m|sz} \right].
\label{eq:yszn_scat}
\ee

We follow the same procedure as in paper II to estimate the $68\%$ confidence region of 
the $\Ysz$--$\Nt$ scaling relation parameters: we randomly sample the $M$--$\Nt$ and $\Ysz$--$\Nt$ parameters
from the appropriate priors, and use the above equations to compute the distribution of the corresponding 
$\Ysz$--$\Nt$ parameters.
The predicted scaling relations for each of the three data sets under consideration --- V09, M10, and P11(z=0.23) --- are summarized in 
Table \ref{tab:rel1}.  

As this work was being completed, \citet{nohcohn12} published another study on observable covariance.  From their table 2,
we compute a correlation coefficient $r_{sz,n|m}\approx 0.4$ in reasonable agreement with the \citet{anguloetal12} value,
and possibly at odds with our interpretation of the latter value as due primarily to aperture-induced covariance.  
If the covariance in the cluster observables if fully dominated by the local cluster triaxiality and/or local filamentary
structure \citep[see e.g.][]{whiteetal10,nohcohn11}, then the good agreement between the \citet{anguloetal12} 
and \citet{nohcohn12} would be expected.
Regardless of the origin of the covariance, we emphasize that our comparison to the maxBCG data explicitly incorporates
the impact of such covariance as described above.


\begin{figure}
\begin{center}
\scalebox{1.2}{\plotone{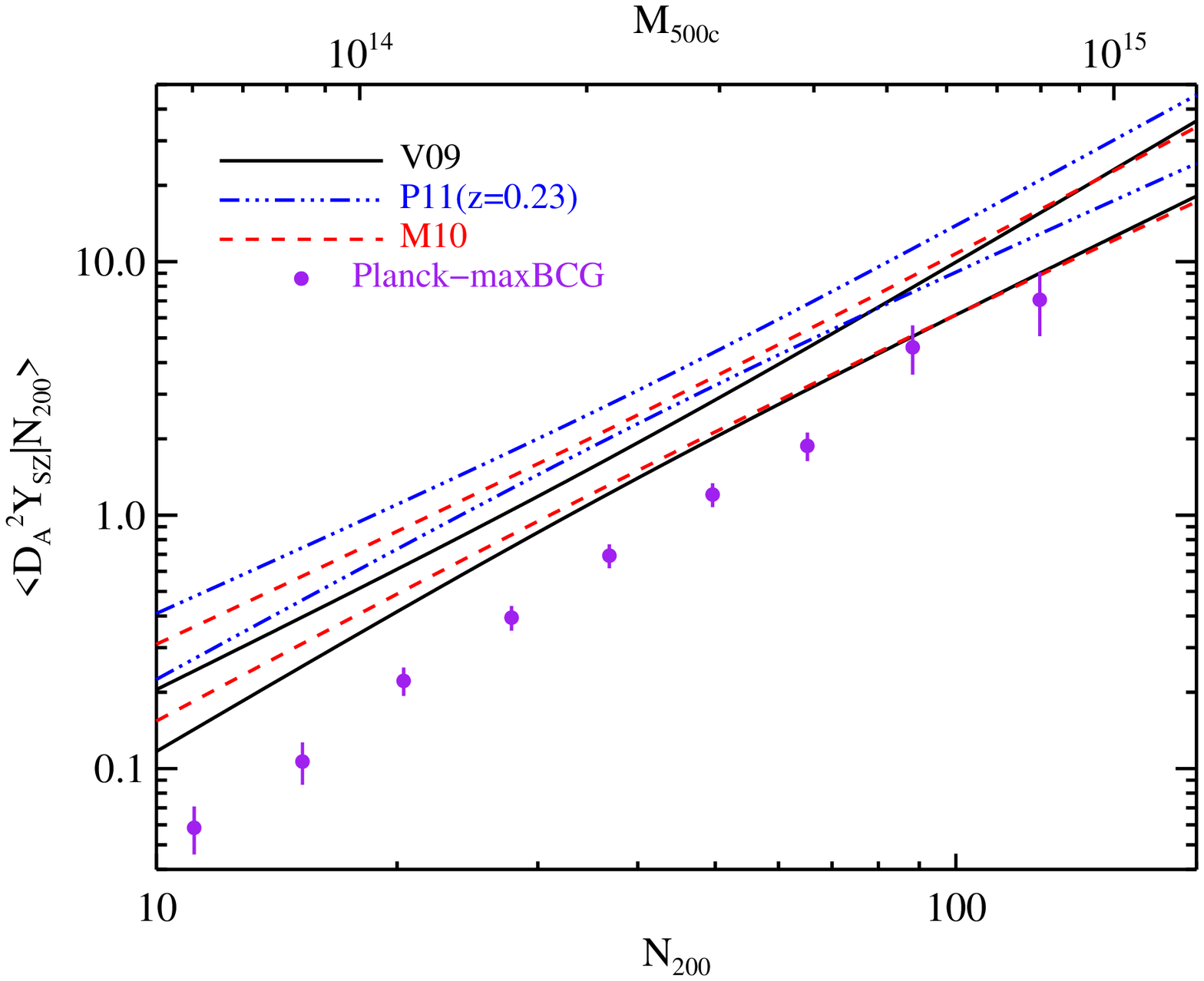}}
\scalebox{1.2}{\plotone{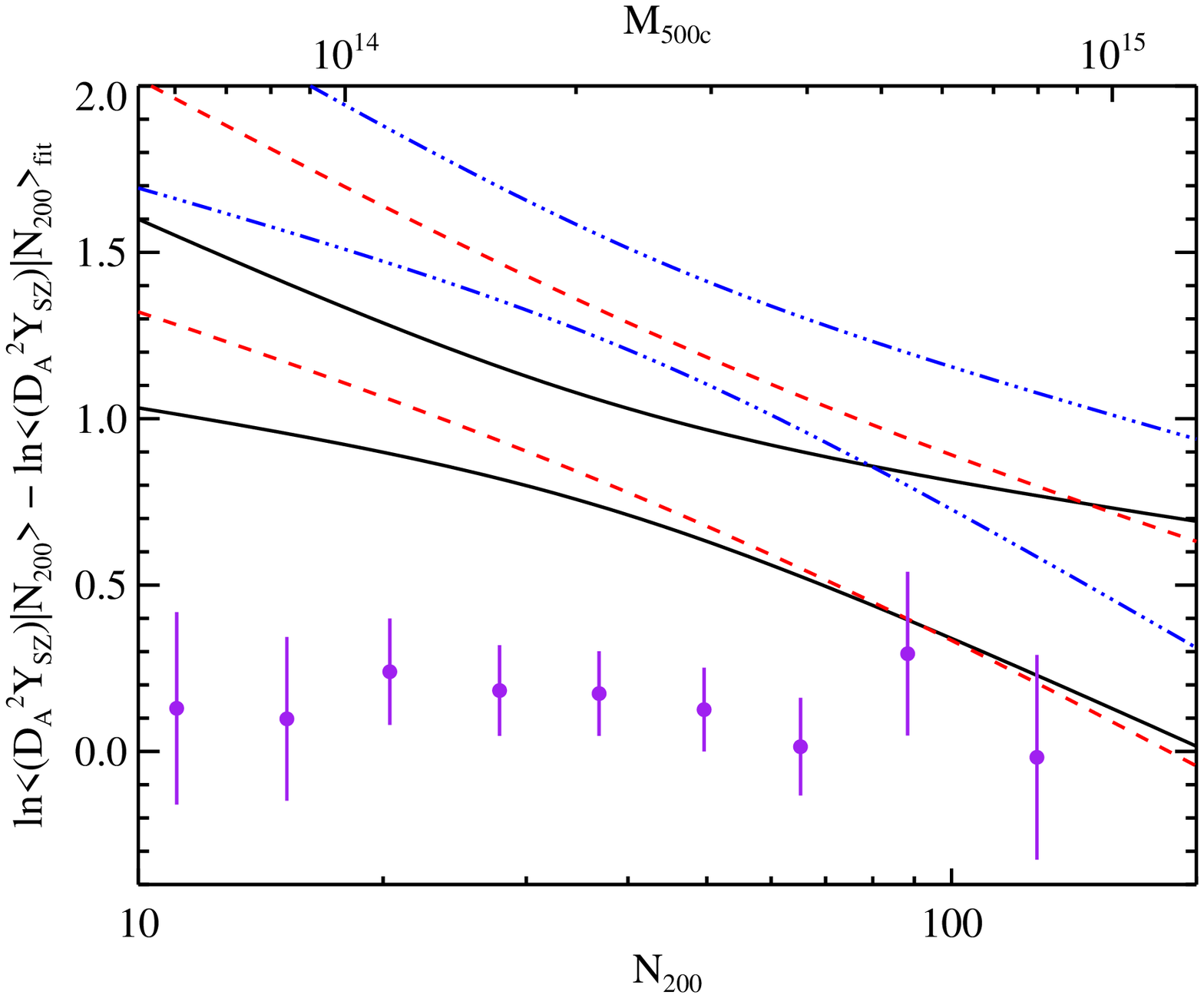}}
\caption{{\it Top panel: } Comparison of the predicted $\avg{\Ysz|\Nt}$ relation for
each of the data sets considered in this work (V09, M10, P11) 
to the \citet[P11-opt,][]{planck11_optical} data, as labelled.  All bands cover the 68\%
confidence regions for each individual scaling relation.
The miscentering corrections of \citet{biesiadzinskietal12} have been applied
on the data drawn from P11-opt.  We have also applied the aperture-induced
Malmquist bias corrections (see text).
 {\it Bottom Panel:}
As top panel, but after subtracting the $\avg{\Ysz|\Nt}$ fit from 
P11-opt.  The data points do not scatter about zero because of
the corrections applied.
We emphasize that the uncertainty in the model predictions are strongly correlated, 
so $\chi^2$-by-eye is grossly misleading.  For instance, lowering the predicted amplitude
for the V09 model by $\approx 3\sigma$ results in good agreement with the data, reflecting
the fact that the tension between the P11-opt data and the V09 model is only $2.8\sigma$.
\vspace{-0.2in}
}
\label{fig:yszn}
\end{center}
\end{figure}



\subsection{Results}

The top panel in Figure \ref{fig:yszn} compares the $\Ysz$--$\Nt$ relation predicted using the V09, M10, and P11(z=0.23)
scaling relations to the P11-opt data.  
Note in Figure \ref{fig:yszn} we plot $\ln \avg{D_A^2\Ysz|\Nt}$ rather than
$\avg{\ln D_A^2\Ysz|\Nt}$ to match the P11-opt data. 
 The bottom panel shows this same data after subtracting the
$\avg{\Ysz|\Nt}$ fit reported in P11-opt (the points do not scatter about zero because of the miscentering and aperture-induced bias corrections applied).   There is reasonable agreement between the \planck--maxBCG data and the V09 and M10 models  at high masses ($M \gtrsim 5\times 10^{14}\ \msun$), 
but all scaling relation predictions fail at low masses.  
Note that because the pivot point of the predicted scaling relations is $\Nt=70$, the
offset shown in Figure \ref{fig:yszn} is not simply a difference in slope:  there is a significant difference in amplitude
as well.   

We emphasize that the uncertainty in our model predictions are comparable to or larger than the errors in the P11-opt data. 
Moreover, these theoretical uncertainties 
are very strongly correlated, so $\chi^2$-by-eye is grossly misleading.  For instance, reducing
the amplitude of our V09 model prediction by $3\sigma$ results in good agreement between the P11-opt data
and our model prediction.  Indeed, the tension between the V09+R09 prediction and the P11-opt data is just
under $3\sigma$.  

We make this quantitative by computing $\chi^2$, adding in quadrature the observational errors
from P11-opt to the covariance matrix for the V09 predictions.  
We evaluate goodness of fit by generating $10^5$ Monte Carlo realizations of the data using the 
full covariance matrix, and
empirically compute $P(\chi^2 \geq \chi^2_{obs})$.  The corresponding probabilities for each of the three
models we consider are $P=0.005$ ($2.8\sigma$, V09), $P\leq 10^{-5}$ ($\gtrsim 4.4\sigma$,P11(z=0.23)),
and $P=0.06$ ($1.9\sigma$, M10).
Evidently, the tension is not anywhere nearly as strong as it appears by eye, though the P11(z=0.23) model
is ruled out at high significance.  If we used only the diagonal terms in the full covariance matrix,
we find $\chi^2_{diag}=71$, which demonstrates how grossly misleading ``$\chi^2$-by-eye'' estimates can be.

Reconciling the $\Ysz$--$\Nt$ data with our predictions requires that at least one of the input scaling relations that were used
in our predictions be incorrect.  
R09 allows for a $10\%$ systematic error offset in the weak lensing masses used to construct the $M$--$\Nt$ relation,
but lowering the cluster masses by said amount is not sufficient to remove the tension from the data.  
If R09 have not underestimated their systematic
uncertainty, the remaining discrepancy would have to be in the X-ray mass estimates.  
Because all X-ray works relied on hydrostatic mass calibration, the most obvious source of bias is 
non-thermal pressure support in galaxy clusters.  We consider a fiducial model in which there is $15\%$ hydrostatic 
bias, meaning hydrostatic masses are $15\%$ lower than the true cluster masses {\it when compared at a fixed aperture.}  
This value is typical of what is predicted from numerical
simulations \citep[$\approx 10\%-25\%$, e.g.][]{nagaietal07a,lauetal09,battagliaetal11,nelsonetal11,rasiaetal12}.

Because cluster masses are defined using a constant overdensity criteria, 
the bias between the true and reported $M_{500}$ values
is larger than the mass bias at fixed aperture.  
For an NFW \citep{nfw96} profile with $\Rf/R_s=3$ \citep[e.g.][]{vikhlininetal06}, we find $M(R)\propto R^{0.88}$ at $R=\Rf$.
Assuming $M_{obs}(R) = bM_{true}(R)$, and expanding in a power-law around $\Rf$, we find that the bias
in the reported $M_{500}$ mass is $M_{500}^{obs}/M_{500}^{true} = b^{1.4}$, or 21\% in our case.
This, then, is the total mass bias we ascribe to the X-ray masses.

\begin{figure*}
\begin{center}
\scalebox{1.2}{\plotone{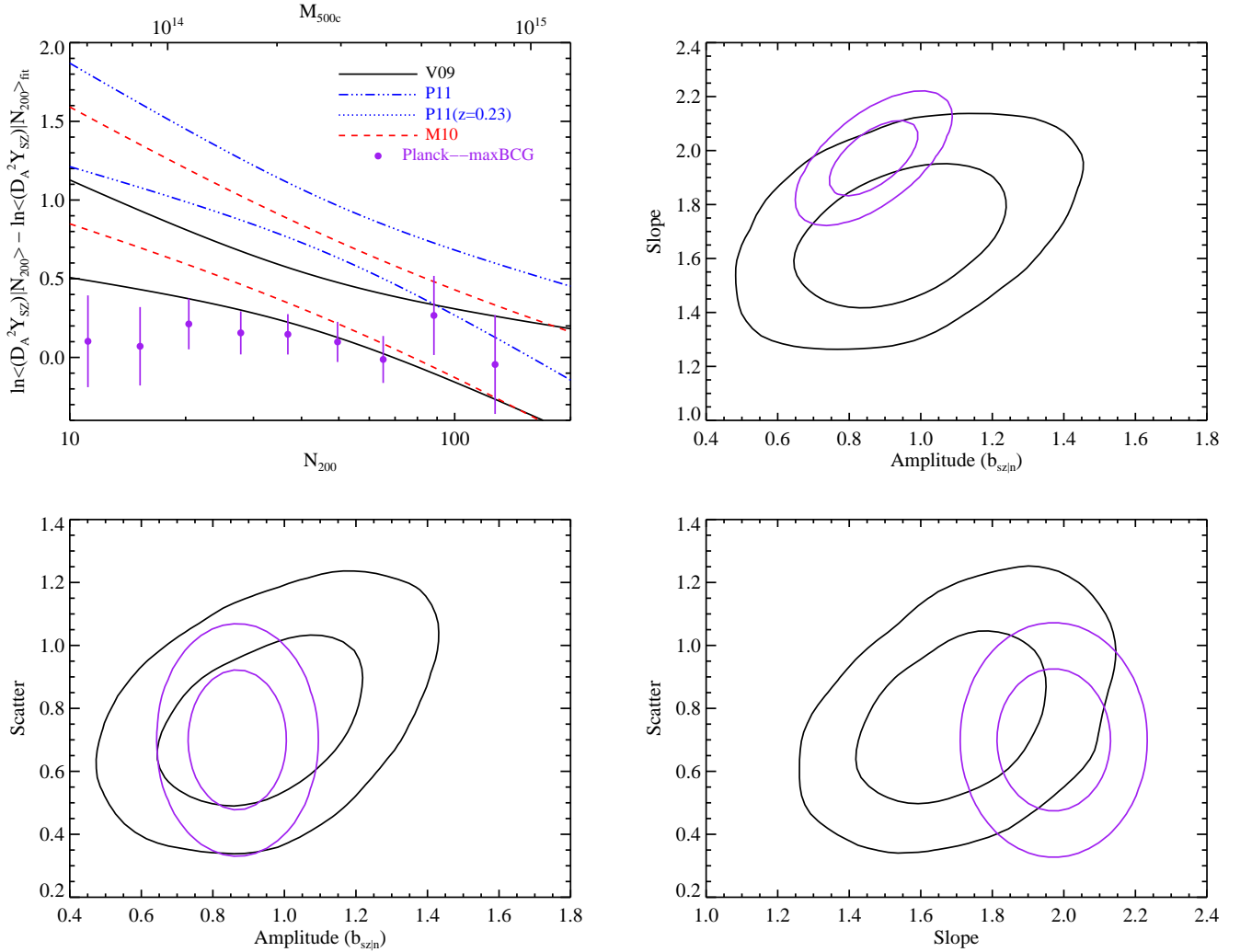}}
\caption{{\it Top Left:} Comparison of the predicted $\avg{\Ysz|\Nt}$ relations 
to the P11-opt data (corrected for miscentering and aperture--induced Malmquist bias).  The prediction assumes a $15\%$ hydrostatic
bias (at fixed aperture, $21\%$ total bias) in X-ray masses, and reduced the \citet{rozoetal09a} mass normalization
by its allotted systematic uncertainty (10\%).   With these corrections, the V09 and M10 models
are in good agreement with the data.  {\it Remaining panels: } 68\% and 95\% confidence contours for the 
observed (purple curves) and predicted (solid curves) $\Ysz$--$\Nt$ scaling relation parameters. The
$\Ysz$--$\Nt$ prediction focuses on the V09 model only.
\vspace{-0.1in}
}
\label{fig:yszn_sys}
\end{center}
\end{figure*}

The top-left panel in Figure \ref{fig:yszn_sys} compares the predicted scaling relations with the P11-opt data
after including these two corrections, i.e. lowering the masses of R09 by 10\% while raising X-ray masses by 21\%.
Possible origins of the 10\% downwards shift of the maxBCG masses relative to R09 will be discussed in 
section \ref{sec:discussion}.  For the time being, we just wish to investigate whether the scaling relations modified in 
this way provide a good description of the data.
As shown in Figure \ref{fig:yszn_sys}, the V09 and M10 models result in excellent fits to the data, with
$P(\chi^2\geq \chi^2_{obs})=0.83$ and $0.52$ respectively.  
The P11(z=0.23) model has
$P(\chi^2\geq \chi^2_{obs})=0.01$, and remains in tension with the data at the $2.5\sigma$ level.  

We have also considered what happens if we increase the bias in either the weak lensing
masses or the X-ray masses.  In either case, increasing the mass bias by up to $\approx 20\%$ results
in good agreement between the predicted and observed $\Ysz$--$\Nt$ relations.
Note, however, that the absence of any is ruled out.
Consequently, we can think of the P11-opt as placing a lower
limit to the hydrostatic and weak lensing bias in the data, while placing only a very weak upper limit on these biases.


We again emphasize that $\chi^2$-by-eye can be
very misleading, and that a better sense of the agreement or tension between the P11-opt data and our
predictions can be obtained by comparing the observed amplitude, slope, and scatter of the $\Ysz$--$\Nt$
relation to our predictions.    This comparison is shown in
the remaining panels of Figure \ref{fig:yszn_sys}.
We focus on the V09 measurements alone to avoid overcrowding the figure.
We take $\sigma_{sz|n}=0.7\pm 0.15$ as the scatter value from P11-opt, which 
is broadly consistent with their Figure 4 across a large richness range.
For this comparison,
we have refitted the P11-opt data after miscentering corrections.  We note 
that the amplitude parameter
plotted is $\tilde a_{sz|n}$ rather than $a_{sz|n}$, so as to match the P11-opt data.  There is good agreement between
the predicted amplitude, slope, and scatter of the $\Ysz$--$\Nt$ relation, with significant overlap between the two distributions.
Because there is covariance between the amplitude and slope of our predicted relation, $\tilde a_{sz|n}$ and $\alpha_{sz|n}$ are
correlated despite the fact that we chose our pivot point to decorrelate $a_{sz|n}$ and $\alpha_{sz|n}$.


\subsection{The MCXC sub-sample of maxBCG Clusters}

One question that we have not yet addressed is that of the SZ signal of the 
MCXC subsample of maxBCG
galaxy clusters.  P11-opt find that when they stack the sub-sample of maxBCG
galaxy clusters that are also in the MCXC catalog \citep{piffarettietal11} --- a heterogeneous
compilation of X-ray selected cluster catalogs --- then the
observed $\Ysz$ signal is boosted by a large amount, and appears to be in reasonable
agreement with the P11-opt predicted scaling relation.   The fact that the SZ signal for the maxBCG--MCXC
cluster subsample is higher than that of the full maxBCG sample has already been quantitatively addressed
by \citet{biesiadzinskietal12} and \citet{anguloetal12}.   Consequently, we do not see
a need for us to repeat their calculations here. 

We would like, however, to present a simple qualitative argument that addresses these
results.  As noted in \citet{rozoetal09a}, the scatter in mass and $\Lx$ at fixed richness are very
strongly correlated ($r\geq 0.85$ at 95\% CL).  This reflects that fact that the maxBCG
richness estimator is very noisy, much noisier than $\Lx$.  Consequently, at fixed richness,
a brighter cluster is always more massive.  By the same token, a brighter X-ray cluster will
also have a higher $\Ysz$ signal.  Given a richness bin, a selection of the X-ray brightest
clusters in the bin necessarily ``peels off'' the high SZ tail of the clusters in the bin.  
Hence, the SZ signal of the X-ray bright maxBCG
sub-sample is higher than that of the full sample, as observed.  Moreover, as one goes lower and lower in richness,
a given luminosity cut will peel off systems
that are further out in the tail of the mass distribution.  That is, the lower the richness bin, the stronger the
impact that the X-ray selection has on the recovered signal, in agreement with the P11-opt data.

Thus, we believe the apparent agreement between the prediction in P11-opt and the MCXC--maxBCG
sub-sample of galaxy clusters is entirely fortuitous.  It is clear that the strong covariance in $\Lx$ and
$\Ysz$ at fixed richness predicts an increase in the SZ amplitude, and a flattening of the slope for
the X-ray bright subsample.  Since the full sample starts below the P11-opt prediction, the X-ray selection
necessarily improves agreement with the model. 
This, however, is a selection effect that happens to move us towards the model prediction, rather than 
the result of a self-consistent model of cluster scaling relations.    Indeed, the P11-opt prediction is
explicitly a prediction for the full maxBCG sample, not for an X-ray bright sub-sample.  Once one accounts
for selection effects, the model curve for the MCXC--maxBCG cluster sub-sample in P11-opt would move
upwards by the same amount that the P11-opt data did, preserving the original tension.

In short, the big puzzle presented by the P11-opt data is not the apparent agreement between the P11-opt
model and the MCXC--maxBCG cluster sample --- that is simply a fortuitous coincidence--- the question is why did
the model fail in the first place.


\subsection{The Bigger Picture}
\label{sec:bigpic}

From Figure \ref{fig:yszn_sys}, we see that the V09 model comes closest to the P11-obs data.  
In the interest of brevity, we will henceforth focus exclusively on the V09 model.  
The results  from papers I and II allow us to quickly infer how the M10 and P11(z=0.23) models
will behave: M10 generically agrees with V09 at high masses, but extrapolates
poorly to low masses due to the constant $\fgas$ assumption, while the P11(z=0.23) masses are biased low 
by $\approx 20\%$ relative to V09 across the observed range.

Having restricted ourselves to the V09 model,
we have shown that a 10\% overestimate of the optical masses, and a 15\% hydrostatic bias (at fixed aperture,
21\% total bias) results in excellent agreement with the data.  To take this solution seriously, however, 
these mass calibration offsets
need not only explain the $\Ysz$--$\Nt$ data, they must also be able to fit all
multi-wavelength data available.  

For the remainder of this work, we consider whether these bias-corrected V09 and R09 scaling relations can satisfy this requirement.
More specifically, for our model to 
be successful it must satisfy the following conditions:
\begin{enumerate}
\item Both optical and X-ray cluster spatial abundances must be consistent
with cosmological expectations in a WMAP7 cosmology.  
\item Optical and X-ray spatial abundances must be consistent with each other.
\item The model must fit all scaling relation data, from both optical, X-ray, and SZ selected cluster catalogs. 
\item We must be able to self-consistently use two observed scaling relations
to predict a third, and the predictions must agree with observations.
\end{enumerate}
We now turn to address each of these points in turn.


\section{spatial abundance Constraints}
\label{sec:spatial abundance}

In paper II, we considered the cosmological consistency of the V09 $\Lx$--$M$ scaling relation. 
Specifically, we showed that by convolving the $\Lx$--$M$ relation of V09 with the halo
mass function  \citep{tinkeretal08} in a WMAP7+BAO+$H_0$ \citep{komatsuetal11}
or a WMAP7+BOSS \citep{sanchezetal12} cosmology,
we can succesfully reproduce the observed REFLEX luminosity function \citep{bohringeretal02}.
An obvious question that arises is whether the bias-corrected V09 scaling relation
is still consistent with cosmological expectations.
Moreover, one must also demand that the spatial abundance of X-ray galaxy clusters
be consistent with the spatial abundance of optical galaxy clusters.  That is, 
convolving $n(\Nt)$ with $P(L_X|\Nt)$ one should recover the X-ray luminosity function.
We now test both of these conditions.

To estimate the X-ray luminosity function from maxBCG data, we
randomly sample the $\Lx$--$\Nt$ relation parameters (including scatter) from \citet{rozoetal09a}, and use these to randomly
assign an X-ray luminosity to every cluster.  We then construct the cumulative
luminosity function $n(\Lx)$ by dividing the recovered spatial abundance by the volume sampled
by the maxBCG catalog.  The whole procedure is iterated $10^4$ times, and we compute
the average cumulative luminosity function along with the corresponding uncertainty, defined as the standard
deviation of our Monte Carlo analysis.  We also correct the luminosity function for the expected evolution
between $z=0.08$, the median redshift of the REFLEX sample, and $z=0.23$, the median redshift of the maxBCG
sample, though we find this evolution is negligible for our purposes.


\begin{figure}
\begin{center}
\scalebox{1.2}{\plotone{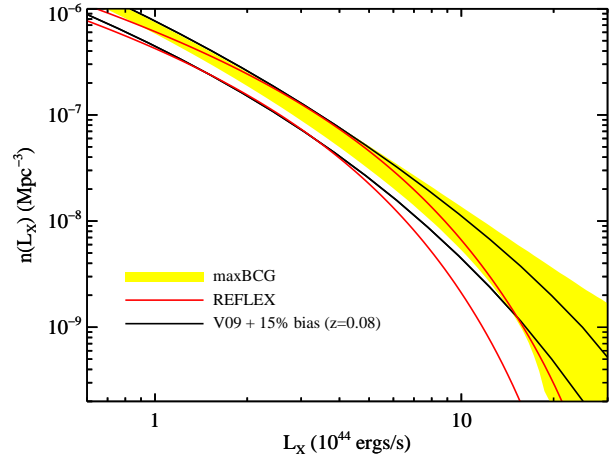}}
\caption{Comparison of the X-ray luminosity function predicted from the maxBCG data (yellow band) to the
REFLEX luminosity function (red band) \citep{bohringeretal02}, and the prediction obtained by convolving the
\citet{tinkeretal08} mass function and the bias-corrected V09 $\Lx$--$M$ scaling relation (black band).
That is, we assume hydrostatic X-ray masses suffer from a 15\% bias at fixed aperture (21\% total bias).
The cosmological model for the V09 prediction is a flat $\Lambda$CDM cosmology as constrained by 
WMAP7+BOSS \citep{sanchezetal12}.  We emphasize that the yellow band from maxBCG data is completely
independent of mass calibration uncertainties, and relies only on the maxBCG abundance and the $L_X$--$\Nt$
scaling relation.
}
\label{fig:lx_lumfunc}
\end{center}
\end{figure}


Figure \ref{fig:lx_lumfunc} shows the maxBCG luminosity function 
as a yellow band.  Also shown are the REFLEX luminosity function (red lines)
and the V09 prediction (black solid lines) for $z=0.08$ using the bias-corrected scaling relation and the
WMAP7+BOSS cosmological constraint
$\sigma_8\Omega_m^{1/2}=0.441\pm 0.013$ \citep[Sanchez, private communication, based on][]{sanchezetal12}.
All three luminosity functions are in excellent agreement, with all relative offsets being significant at less than $0.3\sigma$.   
In principle, we should sample all cosmological parameters, but because the full covariance matrix of the cosmological
parameters in \citet{sanchezetal12} is not yet publicly available, we limited ourselves to sampling only the combination
$\sigma_8\Omega_m^{1/2}$.   We note, however, that when
using the WMAP7+BAO+$H_0$ chains of \citet{komatsuetal11}, we find that varying all
cosmological parameters or only $\sigma_8\Omega_m^{1/2}$ yields very similar results, 
as expected \citep[see e.g. the discussions in ][]{weinbergetal12}.

We can also directly test whether the maxBCG function and the $M$--$\Nt$ relation from R09
are consistent with cosmological expectations.  We
randomly draw the parameters of the $M$--$\Nt$ relations from Table \ref{tab:rel1}, and use the resulting
$P(M|\Nt)$ distribution to randomly assign masses to each maxBCG galaxy cluster. 
Knowing the volume sampled by the maxBCG systems ($z\in[0.1,0.3]$, $7398\ \deg^2$), 
we compute the corresponding 
cumulative mass function, and compare it to 
predictions from the \citet{tinkeretal08} mass function, sampled over the cosmological constraints from
\citet[][]{sanchezetal12}.  As above, we sample only $\sigma_8\Omega_M^{1/2}$, holding the remaining
cosmological parameters fixed.
The whole procedure is iterated $10^3$ times, and the mean and variance of the difference
between the two mass functions is stored
along a grid of masses.  The 68\% confidence band is estimated using the $\pm\sigma$
region where $\sigma$ is the measured standard deviation.


\begin{figure}
\begin{center}
\scalebox{1.2}{\plotone{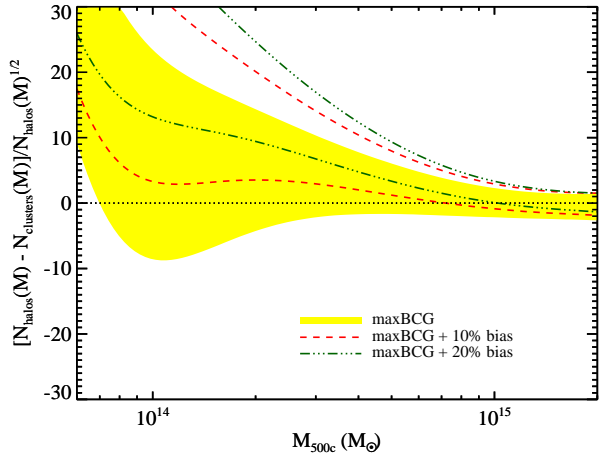}}
\caption{Comparison of the mass function predicted from the maxBCG data to the \citet{tinkeretal08}
mass function for a flat $\Lambda$CDM cosmology as constrained by WMAP+BOSS \citep{sanchezetal12}.
Lowering the maxBCG masses by 10\% is consistent with cosmological expectations.  A 20\% shift, however, 
starts to be in mild ($2\sigma$) tension with the data.  Thus, we take this as roughly the maximum correction compatible with the
\citet{sanchezetal12} results.
}
\label{fig:mf}
\end{center}
\end{figure}


Figure \ref{fig:mf} shows the difference between the cluster and halo mass functions, in units
of the Poisson error.  The statistical error in the spatial abundance is super-Poisson \citep{hukravtsov03},
but the uncertainty in the difference is actually dominated by uncertainties in cluster masses rather
than cluster statistics \citep[e.g][]{weinbergetal12}.
The yellow band in Figure \ref{fig:mf} shows the $68\%$ confidence contours for our fiducial case.
The fact that this band covers the line $y=0$ demonstrates that the fiducial \citet{rozoetal09a} $M$--$\Nt$ relation and the maxBCG
spatial abundance function are consistent with cosmological expectations.   The up-turn at low
masses marks the scale at which the maxBCG catalog becomes incomplete. 

The red dashed lines in Figure \ref{fig:mf} shows the 68\% confidence contours for the 
difference between the cluster and halo mass functions
after reducing the amplitude of the \citet{rozoetal09a}  $M$--$\Nt$ relation by 10\%.    
The difference between the resulting maxBCG data and the theoretical mass functions is $\approx 1.4\sigma$,
so this shift in cluster masses is statistically allowed.
Lowering the amplitude by 20\% --- shown as the green dashed--dotted curves --- results in a marginally unacceptable
fit, with $2.2\sigma$ tension. 
Thus, a $20\%$ bias on the
weak lensing mass scale from R09 is roughly the maximum level of bias allowed 
from a cosmological perspective.  

By the same token, one may ask what level of hydrostatic bias does the X-ray data allow, in the sense that the predicted and observed
X-ray luminosity functions must be consistent with each other.
We find that up to a $\approx 30\%$ hydrostatic bias at fixed aperture ($\approx 40\%$ total) is acceptable ($\lesssim 2\sigma$)
when using WMAP7+BOSS priors.  
These results are entirely consistent with the biases required to explain the $\Ysz$--$\Nt$ relation, i.e. 10\% bias in the maxBCG
masses, and a $15\%$ hydrostatic bias in X-ray masses.
Interestingly, this also demonstrates that cosmological considerations can place an upper limit for the
hydrostatic and weak lensing biases in the data.  In conjunction with the lower limit placed
by the $\Ysz$--$\Nt$ relation, these limits nicely bracket our proposed solution.


\section{Optical and X-ray Consistency}
\label{sec:optical_xray_SZ_consistency}

V09 calibrated the $\Lx$--$M$ scaling relation, as well as the $M$--$Y_X$
scaling relations.  Using the $\Ysz$--$Y_X$ scaling relation from \citet{rozoetal12a}, 
in paper II we used this data to also produce a V09 $\Ysz$--$M$ scaling relation.
We now derive the $\Lx$--$M$ and $\Ysz$--$M$ relations from maxBCG data, and compare them
to the V09 results.

\subsection{Method}

We rely on a Monte Carlo method to compute the scaling relations predicted from maxBCG.
Specifically, we use the $M$--$\Nt$, $\Lx$--$\Nt$, $\Ysz$--$\Nt$ scaling relations to randomly assign
masses, X-ray luminosities, and SZ signals to each maxBCG galaxy cluster.  We then select
a mass-limited cluster sub-sample with $M\ge 2\times 10^{14}\ \msun$ --- comfortably above
the completeness limit of the sample, see Figure \ref{fig:mf} ---  and we fit for the resulting 
scaling relation parameters.  
The whole procedure is iterated $10^4$ times.  Each scaling relation is then evaluated along a grid
in mass, and the mean and standard deviation at each point computed.  The 68\% confidence
intervals are estimated as the $\pm \sigma$ regions where $\sigma$ is the standard deviation.

There is one additional key consideration when predicting the scaling relations from the maxBCG
data: the cluster variables $\Lx$, $\Ysz$, and $M$, are  all tightly correlated with each other at fixed
richness \citep[R09;][]{staneketal10,whiteetal10,anguloetal12,nohcohn12}.  This reflects the fact that $\Nt$ is a very 
poor mass tracer with very large scatter: 
a cluster that is brighter in X-rays will also be more massive and have a higher
SZ signal.  We set the correlation coefficient between the various observables using the
local power-law multi-variate scaling relations model detailed in Appendix A of paper II.
Specifically, the covariance between a quantity $\psi$ and $m$ at fixed richness (subscript $n$)
is given by 
\be
r_{\psi,m|n} = \frac{ \sigma_{m|n}/\sigma_{m|\psi}-r }{\left[ 1 - r^2 + (\sigma_{m|n}/\sigma_{m|\psi} - r)^2 \right]^{1/2} } \label{eq:r}
\ee
where $r$ is the correlation coefficient between $x$ and $n$ at fixed $m$.   We take the simplifying assumption that this intrinsic covariance is zero ($r=0$), which implies $r_{x,m|n} = 0.90$ and $r_{sz,m|n}=0.98$.
Since the mass scatter at fixed optical richness ($\sigma_{m|n}$) is significantly larger its counterpart at fixed X-ray luminosity ($ \sigma_{m|x}$), 
the precise value of $r$ has only a mild impact; varying $r \in [-0.5,0.5]$ varies $r_{x,m|n}$ over the range $r_{x,m|n}\in [0.86,0.93]$,
with even a smaller range of variation for $r_{sz,m|n}$.

Equation~(\ref{eq:r}) specifies the correlation coefficient we employ when assigning cluster properties to the
maxBCG galaxy clusters.  When evaluating the correlation coefficient, we hold the scatter $\sigma_{m|x}$ fixed
to its central value.   This is because the main source of scatter in $r_{x,m|n}$ 
is the uncertainty in $\sigma_{m|n}$.  In addition, it avoids introducing covariance between fluctuations in 
the scaling relation predicted from maxBCG data and the V09 predictions.   Because $\sigma_{m|n}$ is drawn for each Monte 
Carlo realization, each realization has an independent $r_{x,m|n}$ and $r_{sz,m|n}$ estimate.

To compute the 68\% confidence regions for the V09 scaling relations, we rely instead on the
method used in paper II: the scaling relations parameters are randomly sampled $10^5$ times from the
priors in Tables \ref{tab:rel1} and \ref{tab:rel2}.  These are used to compute the scaling relation parameters
using Appendix A in paper II, which we in turn us to evaluate the
mean and standard deviation
of the scaling relations along a grid of masses.   The 68\% confidence intervals are estimated as above.
In all cases, the biased-corrected scaling relations are computed by modifying the amplitude of the bias-free
scaling relations involving mass by the appropriate amounts (10\% for R09, 21\% for V09).  
We also rescale cluster observables defined within $\Rf$ as in
papers I and II to account for the change in aperture due to mass rescaling, though we note these corrections
are typically small relative to the impact of the mass offset.


\begin{deluxetable*}{lllllll}
\tablewidth{0pt}
\tablecaption{Input Scaling Relations at $z=0.23$}
\tablecomment{
Conventions as per Table \ref{tab:rel1}.  X-ray luminosity is measured in the $[0.1,2.4]\ \keV$ band,
in units of $10^{44}\ \mbox{ergs/s}$.   
}
\tablehead{
Relation & $\chi_0$ & $a_{\psi|\chi}$ & $\alpha$ & $\sigma_{\ln \psi|\chi}$ & Sample}
\startdata
$\Lx$--$M$ & 4.8 & $1.16\pm 0.09$ & $1.61\pm 0.14$ & $0.40\pm 0.04$ & V09 \vspace{1.0truept}\\
$D_A^2\Ysz$--$\Nt$ & 40.0 &	$-0.20\pm 0.13$ & $1.95\pm  0.07$ & $0.70\pm 0.15$ & maxBCG \vspace{1.0truept}\\
$\Lx$--$\Nt$ & 40.0 & $0.04\pm 0.04\ (ran) $ & $1.63\pm 0.06\ (ran)$  & $0.83\pm 0.03\ (ran) $ & maxBCG \\
 &  & $~~~~~~ \pm 0.09\ (sys)$ & $~~~~~~ \pm 0.05\ (sys)$	 & $~~~~~~\pm 0.10\ (sys)$ & \vspace{0.03in} 
\enddata
\label{tab:rel2}
\end{deluxetable*}



\subsection{Results}


\begin{figure*}
\begin{center}
\scalebox{1.2}{\plotone{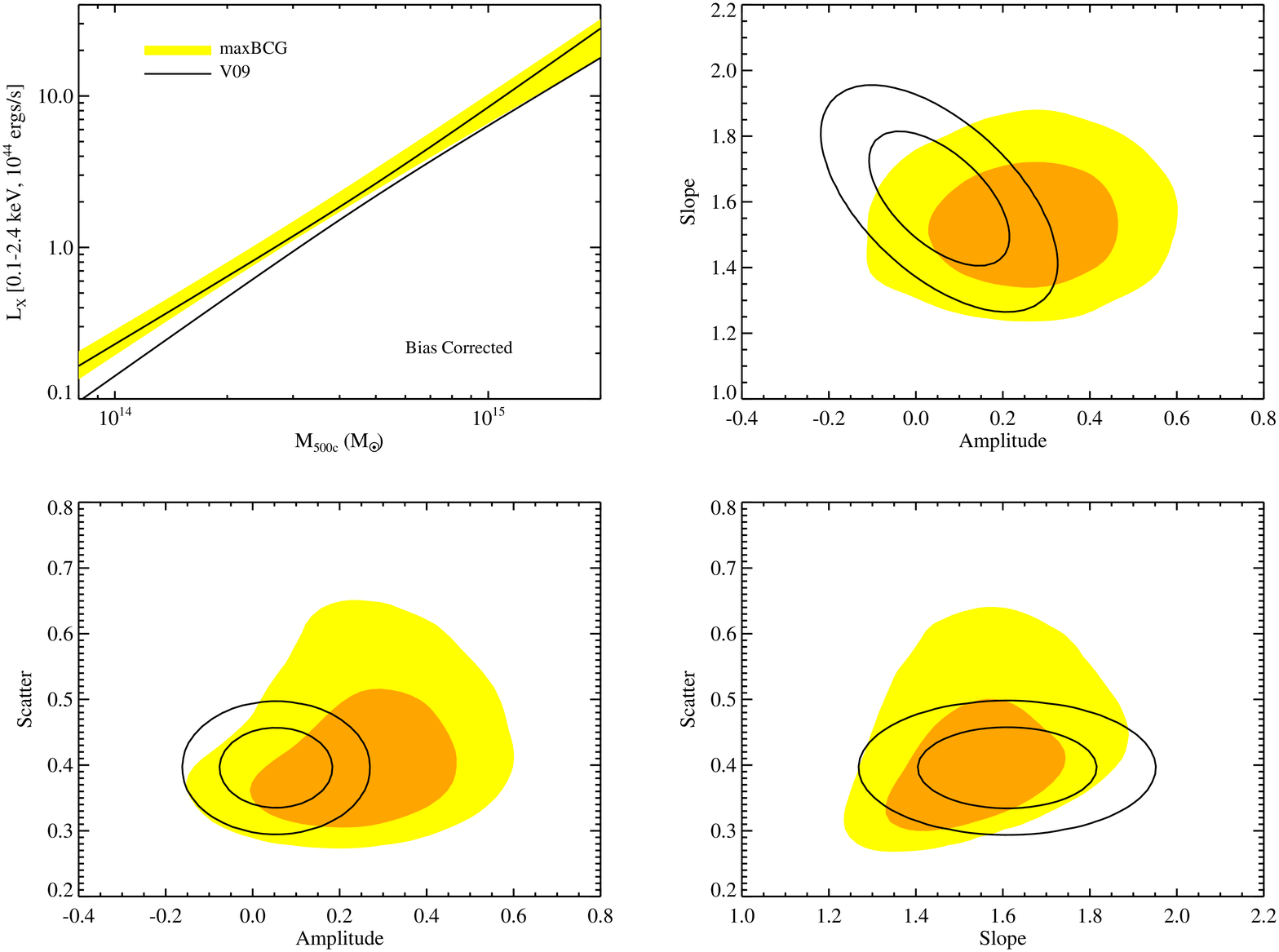}}
\caption{{\it Top-left panel:} 68\% confidence interval for the
$\Lx$--$M$ scaling relation at $z=0.23$ as estimated from the
maxBCG catalog and V09 data sets.  Both sets of scaling relations are bias-corrected, assuming a 10\% overestimate of the
maxBCG masses, and a 15\% hydrostatic bias (at fixed aperture, 21\% total bias) on the X-ray masses.
{\it Remaining panels:} 68\% and 95\% confidence regions of the 
 $\Lx$--$M$ parameter for the bias-corrected
maxBCG (filled contours) and V09 (solid black curves) data sets.  The amplitude parameter is defined at the
pivot point $M=3\times 10^{14}\ \msun$, which is appropriate for the bias-corrected maxBCG data set.
}
\label{fig:lxm}
\end{center}
\end{figure*}


The top-left panel in Figure \ref{fig:lxm} shows the 68\% confidence contours for the V09 and maxBCG 
$\Lx$--$M$ scaling relation.  Both sets of relations are bias-corrected, and are in
excellent agreement with each other.  The scatter, too, is in good agreement, 
with $\sigma_{\ln \Lx|M}=0.38^{+0.10}_{-0.05}$
for the maxBCG prediction, and $\sigma_{\ln \Lx|M}= 0.40\pm 0.04$ for the V09 scaling relation.
This agreement is better illustrated in the remaining panels, where we explicitly compare the predicted
amplitude, slope, and scatter of the $\Lx$--$M$ relation from the two works.  In all cases, the
filled contours correspond to the maxBCG bias-corrected scaling relations, while the solid curves
demarcate the bias-corrected V09 scaling relation. 

\begin{figure*}
\begin{center}
\scalebox{1.2}{\plotone{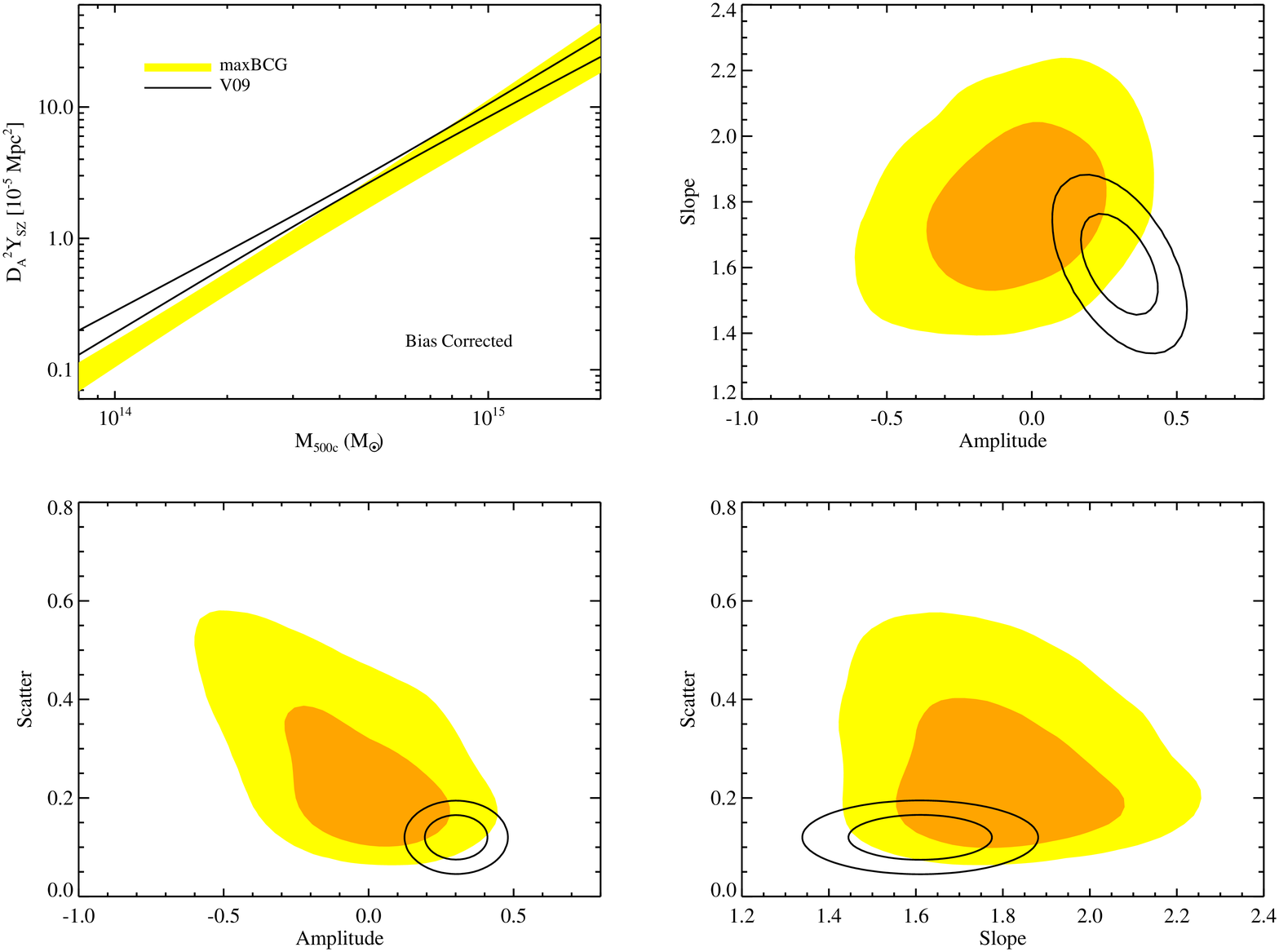}}
\caption{{\it Top-left panel:} 68\% confidence interval for the
$\Ysz$--$M$ scaling relation at $z=0.23$ as estimated from the
maxBCG catalog and V09 data sets.  Both sets of scaling relations are bias-corrected, assuming a 10\% overestimate of the
maxBCG masses, and a 15\% hydrostatic bias (at fixed aperture, 21\% total bias) on the X-ray masses.
{\it Remaining panels:} 68\% and 95\% confidence regions of the 
 $\Ysz$--$M$ parameter for the bias-corrected
maxBCG (filled contours) and V09 (solid black curves) data sets.  The amplitude parameter is defined at the
pivot point $M=3\times 10^{14}\ \msun$, which is appropriate for the bias-corrected maxBCG data set.
}
\label{fig:yszm}
\end{center}
\end{figure*}

Figure \ref{fig:yszm} mirrors Figure \ref{fig:lxm}, only there we are considering the $\Ysz$--$M$
scaling relations.  The top-left panel shows the 68\% confidence interval in the $\Ysz$--$M$
plane, while the remaining panels show the 68\% and 95\% confidence contours for the
scaling relation parameters.  In all cases, we consider only the bias-corrected scaling relations.
Even though the agreement is less good than in the case of the $\Lx$--$M$ relation,
the agreement between the two data sets is still adequate.

Despite the two data sets being in statistical agreement, it is worth asking what would be required
to improve upon the current status, particularly in the case of $\Ysz$--$M$, where the agreement
is less good.  Because the maxBCG band falls below the V09 data in the $\Ysz$--$M$ plane,
but above the V09 data in the $\Lx$--$M$ plane, shifting the cluster masses will increase tension
in one relation while alleviating the other, and is therefore not a good avenue for improving things.
In addition, the X-ray luminosities are tied down by the
spatial abundance constraints from the previous section.  Consequently, any improvement to the current scenario
would have to come from $\approx 10\%-20\%$ changes in the $\Ysz$ measurements: either the V09 $\Ysz$ measurements
would have to go down (e.g. due to Malmquist bias due to SZ cluster selection), or the P11-opt measurements
would have to go up.  Since the V09 $\Ysz/Y_X$ ratio is already on the low end of what is expected,
an increase of the $\Ysz$ signal of maxBCG clusters is the most promising avenue.  

If we repeat this analysis using the original maxBCG and V09 scaling relations, we find that while
the two data sets are in modest agreement with regards to the $\Lx$--$M$ relation, the
$\Ysz$--$M$ relation of the two works are clearly discrepant.  This disagreement reflects the 
original tension in the $\Ysz$--$\Nt$ relation, so it is not surprising that fixing the $\Ysz$--$\Nt$
relation also results in consistent estimates for the $\Ysz$--$M$ relation.  That said, it is far from
trivial that this same solution is consistent with the $\Lx$--$M$ relation, and with the
spatial abundance constraints discussed in the previous section.


\section{Closure Test: the $\Ysz$--$\Lx$ Relation}
\label{sec:self-consistency}

We have seen that the bias-corrected $\Lx$--$M$ and $\Ysz$--$M$ maxBCG and V09 scaling
relations are in good agreement with each other, and are also consistent with cosmological expectations.
We now test whether the direct observable scaling relations are self-consistent.  
Specifically, we test whether the $\Ysz$--$\Lx$ relation predicted from
the $\Ysz$--$\Nt$ and $\Lx$--$\Nt$ relations is consistent with measurements from X-ray cluster catalogs.
Further, we will also compare the predicted $\Ysz$--$\Lx$ scaling relations
to the bias-corrected prediction from the V09 model.

The maxBCG predictions are computed as before: we randomly assigned $\Lx$ and $\Ysz$ values
to clusters in the catalog, including the effect of the correlation coefficient.  
The correlation coefficient between $\Ysz$ and $\Lx$ at fixed richness is estimated
following an argument similar to the previous section.  Specifically, we assume $r_{n,x|sz}=0$,
and estimate
$r_{x,sz|n}$ using the formulae in Appendix A of paper II.  Typical values for $r_{x,sz|n}$
lie in the range $[0.8,1]$, as expected.
We then cut the resulting cluster catalog at $\Lx\geq 8\times 10^{43}\ \mbox{ergs/z}$, which is 
sufficiently high to not be affected by completeness issues, and fit the data to arrive
at our model amplitude, slope, and scatter about the mean.
The V09 predictions are performed exactly as in paper II, where we also considered the $\Ysz$--$\Lx$ relation for the
V09, M10, and P11 data sets, only we now consider the bias-corrected V09 scaling relations only.


\begin{figure*}
\begin{center}
\scalebox{1.2}{\plotone{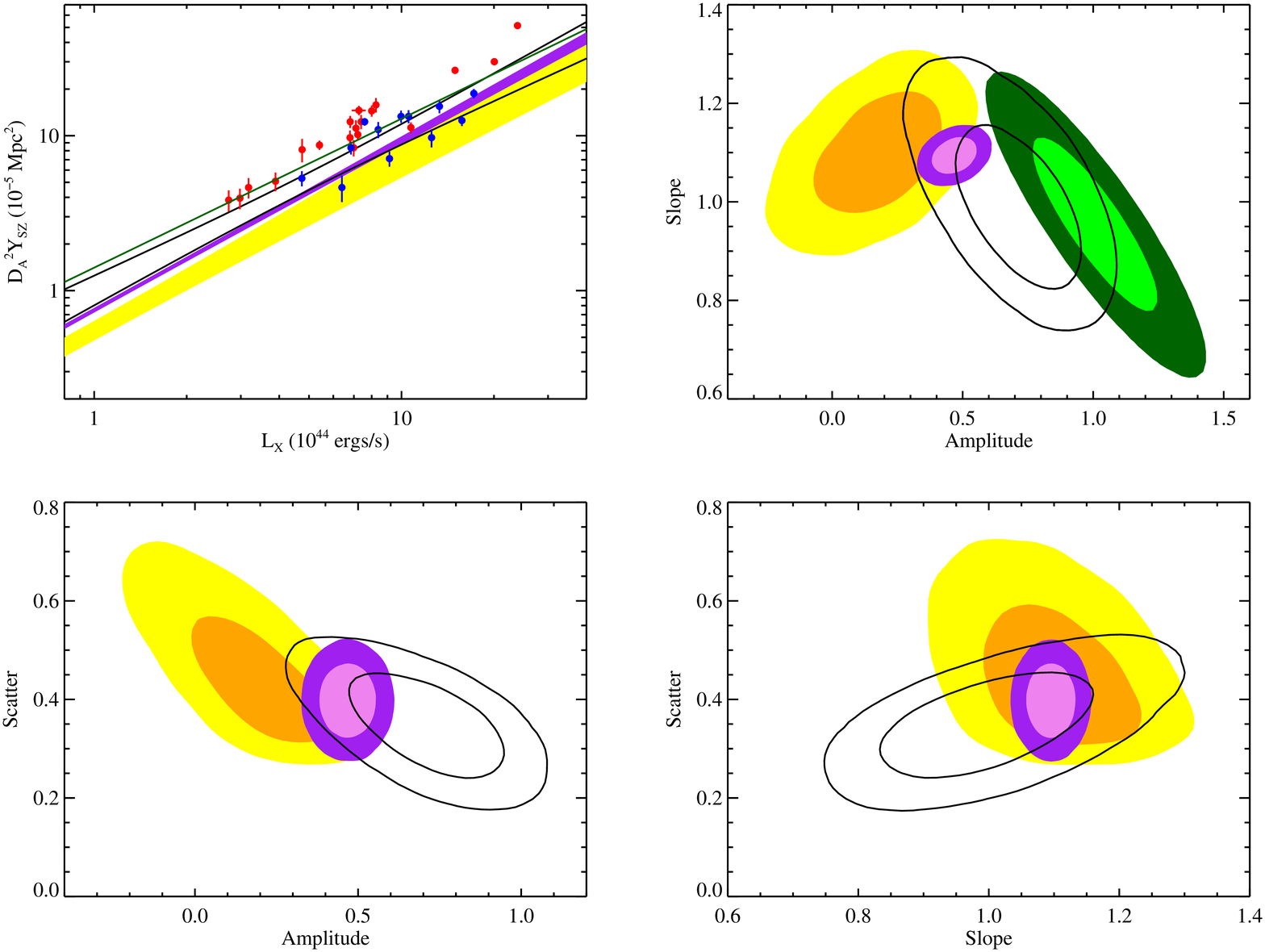}}
\caption{{\it Top-left panel:} 68\% confidence interval for the
$\Ysz$--$M$ scaling relation at $z=0.23$ as estimated from the
maxBCG catalog (yellow band) and V09 (black lines) data sets.  
Both sets of scaling relations are bias-corrected, assuming a 10\% overestimate of the
maxBCG masses, and a 15\% hydrostatic bias (at fixed aperture, 21\% total bias) on the X-ray masses.
Also shown as a purple band is the 68\% confidence interval quoted in \citet{planck11_xray} from
stacking X-ray selected
clusters from the MCXC catalog \citep{piffarettietal11}.  We have corrected these fits for
Malmquist bias due to correlated scatter, and aperture corrections due to the mass bias relative
to V09 (see text).  The green line is our own fit to the P11(z=0.23) data, modified for selection effects as described
in P11.   The data points with error bars
are clusters with $z\in[0.13,0.3]$ in P11.  Blue points are cool-core systems, while red points are not.
The green line is the best fit relation to the P11(z=0.23) data.
{\it Remaining panels:} 68\% and 95\% confidence regions of the 
 $\Ysz$--$M$ parameter for the bias-corrected
maxBCG (yellow contours) and V09 (solid black curves) data sets.  The amplitude parameter is defined at a
pivot point $\Lx=2\times 10^{44}\ \mbox{ergs/s}$, which is a compromise between the V09 and maxBCG
predictions.  The green contours corresponds to our fit of the P11 data, while the purple contours
are those obtained from the MCXC catalog.
}
\label{fig:yszlx}
\end{center}
\end{figure*}


We will be comparing the maxBCG and V09 predictions for the $\Ysz$--$\Lx$ scaling relation 
to those derived from the P11(z=0.23) data, and that obtained by stacking X-ray galaxy clusters from
the MCXC cluster catalog \citep{piffarettietal11} as described in \citet{planck11_xray}, hencerforth
referred to as P11-X.
For the P11(z=0.23) data, we fit the clusters in P11 in the redshift range $z\in[0.13,0.3]$, and modify
the recovered amplitude as described in P11.  We do not attempt to measure the scatter because
of selection effects.  Rather, our fits assume an intrinsic scatter in $\Ysz$--$\Lx$ of $0.4$, which is consistent
with all data.  

As for the P11-X fits to the $\Ysz$--$\Lx$ scaling relation, we perform three independent
corrections: 1) Malmquist bias corrections due two covariance between $\Ysz$ and $\Lx$.  2) Aperture
induced corrections due to the difference in mass calibration between P11 and the bias-corrected V09
model.  3) A correction to account for the fact that P11-X measure $\avg{\Ysz|\Lx}$ rather
than $\avg{\ln \Ysz|\Lx}$.

The first of these corrections --- the Malmquist bias due to correlated scatter ---
works exactly as per our discussion of the $\Ysz$--$\Nt$ scaling relation, with an additional
important consideration: in addition to aperture-induced covariance, $\Ysz$ and $\Lx$ are expected to be
correlated because they probe the same intra-cluster medium.  The correlation coefficients predicted in
\citet{staneketal10} and \citet{anguloetal12} are $r\approx 0.8$ and $r\approx 0.5$ respectively.  
We set $r_{sz,x|m}=0.65\pm 0.2$,
where the uncertainty in $r$ is propagated into the uncertainty of the amplitude of the $\Ysz$--$\Lx$ relation.
We also use the scatter estimates for $\sigma_{sz|m}$ and $\sigma_{x|m}$ from the V09 model, and propagate
the associated uncertainty.  The net effect of this correction is to lower the $\Ysz$ amplitude by $\approx 7\%\pm 2.5\%$.

The aperture correction to the P11-X data has to do with the dependence of $\Ysz$ on the integration
aperture $\Rf$.  From paper 1, we know that a bias $b_m$ in the mass induces an amplitude shift $\Delta a = 0.27\ln b_m$
on $\Ysz$.  The P11(z=0.23) masses are biased by $\Delta \ln M \approx 0.2$ relative to V09, which itself suffers
from a $15\%$ hydrostatic bias in our bias-corrected model.  All told, this increases the $\Ysz$ amplitude by 
$\approx 10\%\pm 3\%$.    

Finally, the cluster stacks in P11-X measure $\avg{\Ysz|\Lx}$ rather than $\avg{\ln \Ysz|\Lx}$, so their reported amplitude must
be lowered accordingly, by $\approx 8\% \pm 2\%$.  Put together, these three corrections result in an
amplitude shift of $\Delta a_{sz|x} = -0.050\pm 0.044$.
We apply this systematic correction to the P11-X data, and add the associated uncertainty in quadrature to their
quoted errors.

The top-left panel in Figure \ref{fig:yszlx} compares the predicted $\Ysz$--$\Lx$ scaling relation derived 
from the V09 data set (black solid lines) with that of the maxBCG catalog (yellow band).  Note for both of these,
we only consider the bias-corrected scaling relations (mass enters via the aperture corrections to $\Ysz$).
The purple band is the observed
relation in the MCXC cluster catalog from P11-X.
The best fit line to the P11(z=0.23) data set is shown as a solid green line.  The points with error
bars are individual clusters in the P11(z=0.23) data set, blue for cool-core systems, and
red for non cool-core systems.

The agreement between the various data sets is best judged using the remaining panels of Figure \ref{fig:yszlx},
where we show the corresponding 68\% and 95\% confidence contours for the scaling relation parameters.  
Focusing on the top-right plot, which shows the amplitude and slope parameters,
we see that the P11(z=0.23) data is in some tension with the other data sets: the amplitude is too high. 
The simplest explanation for
this offset would be an underestimate of the SZ selection effects in P11.

There is, however, reasonable agreement between the maxBCG, V09, and P11-X data sets:
the $1\sigma$ P11-X data curve in the top-right panel 
overlaps the $1\sigma$ contours for both the maxBCG and V09 data sets.
All three data sets are in excellent agreement on the intrinsic scatter in the $\Ysz$--$\Lx$ relation.
To achieve better agreement,  the $\Ysz$ signal of the maxBCG galaxy clusters would need to
increase by $\approx 10\%-20\%$, especially at low masses; a decrease of the V09 $\Ysz$ signal would be difficult to reconcile
with X-ray expectations for the $\Ysz/Y_X$ ratio, while shifts of the X-ray luminosity are limited by
spatial abundance constraints.


\section{The Mass Scale of Galaxy Clusters}
\label{sec:maxbcg_masses}

\subsection{The Mass Scale of maxBCG Galaxy Clusters}

We have already argued that the origin of the bias in X-ray masses proposed in this paper is simple hydrostatic bias.
One interesting question that remains open, however, is where did the bias in the stacked weak lensing mass calibration in R09
come from?
At least part of this answer has to do with the impact of covariance between mass and weak lensing
mass at fixed richness.  Covariance between $\Mwl$ and $\Nt$ introduces corrections in mass calibration
that were ignored in \citet{johnstonetal07} and R09.  Specifically,
from Appendix A in paper II we have
\bea
\avg{ \ln \Mwl|\Nt} & = & a_{wl|m} + \alpha_{wl|m}\avg{\ln M |\Nt} \nn \\
	& & \hspace{0.2in} + r_{wl,n|m}\beta \alpha_{wl|m} \sigma_{m|wl}\sigma_{m|n} 
\eea
where we have assumed
\be
\avg{\ln \Mwl |M} = a_{wl|m} + \alpha_{wl|m}\ln M.
\ee
Assuming the weak lensing mass estimates are unbiased in the sense that
$\avg{\ln \Mwl|M}=\ln M$, we can set $a_{wl|m}=0$
and $\alpha_{wl|m}=1$.\footnote{In practice, unbiased weak lensing masses would result in $\avg{\Mwl|M}=M$ rather than $\avg{\ln \Mwl|M}=\ln M$. 
The difference between these two assumptions is a net offset of $0.5\sigma_{wl|m}^2 \approx 2\%$, which is negligible for our purposes.}
Inserting this into our expression for $\avg{\ln \Mwl|\Nt}$, and solving for $\avg{\ln M|\Nt}$
we find
\vspace{0.05in}
\be
\avg{\ln M |\Nt} = \avg{\ln \Mwl|\Nt} - r_{wl,n|m}\beta\sigma_{m|wl}\sigma_{m|n}
\ee


\begin{deluxetable*}{lllclll}
\tablewidth{0pt}
\tablecaption{Preferred Set of Scaling Relations}
\tablecomment{
X-ray luminosity is measured in the $[0.1,2.4]\ \keV$ band
in units of $10^{44}\ \mbox{ergs/s}$.  $D_A^2\Ysz$ is in units of $10^{-5}\ \Mpc^2$.
The maxBCG scaling relations are bias-corrected, while the
V09+maxBCG relations are the joint constraint from the bias-corrected V09 and maxBCG samples.
Scaling relations involving mass include a $\pm 10\%$ systematic uncertainty in the mass. 
The error in the amplitude of the $\Ysz$--$\Lx$ relation is larger than that quoted in 
P11-X because we include the uncertainty in our systematic corrections.
This set of scaling relations is fully self-consistent.
}
\tablehead{
Relation & $\chi_0$ & Amplitude ($a_{\psi|\chi}$) & $\alpha_{\psi|\chi}$ & $\sigma_{\ln \psi|\chi}$ & Sample}
\startdata
$\Lx$--$M$ & 4.4 & $0.72\pm 0.07\ (ran) \pm 0.16\ (sys)$ & $1.55\pm 0.09$ & $0.39\pm 0.03$ & V09+maxBCG      \\
$D_A^2\Ysz$--$M$ & 4.4 & $0.87 \pm 0.06\ (ran) \pm 0.17\ (sys)$ & $1.71\pm 0.08$ & $0.15\pm 0.02$ & V09+maxBCG 
\vspace{0.05in} \\
\hline 
\hline \vspace{-0.05in} \\
$M$--$\Nt$ & 40 & $0.75\pm 0.10$ & $1.06\pm 0.11$ & $0.45\pm 0.10$ & maxBCG \\
$\Lx$--$\Nt$ & 40 & $0.04\pm 0.10$ & $1.63\pm 0.08$ & $0.83\pm 0.10$ & maxBCG \\
$\Ysz$--$\Nt$ & 40 & $-0.24\pm 0.20$ & $1.97 \pm 0.10$ & $0.70 \pm 0.15$ & maxBCG
\vspace{0.05in} \\
\hline 
\hline \vspace{-0.05in} \\
$\Ysz$--$\Lx$ & 1.0 & $-0.29\pm 0.06$ & $1.10\pm 0.03$ & $0.40\pm 0.05$ & P11-X
\vspace{0.05in}
\enddata
\label{tab:final_rel}
\end{deluxetable*}


We see then that any covariance between $\Mwl$ and $\Nt$ at fixed mass implies that the weak lensing masses 
must be corrected downwards.  This is exactly the same type of Malmquist-bias correction that we applied on
the $\Ysz$--$\Nt$ and $\Ysz$--$\Lx$ scaling relations, though this covariance is not aperture-induced.
In \citet{anguloetal12}, the correlation coefficient $r_{wl,n|m}$ is estimated to be $r\approx 0.4$, in good
agreement with the value from \citet{nohcohn12}.
Assuming $\sigma_{wl|m}\approx 0.2$ as per \citet{beckerkravtsov10}, $\sigma_{n|m}\approx 0.4$, and $\beta\approx 3$,
and setting $r$ to the \citet{anguloetal12} value, we find that the mass calibration from R09 should be corrected
downwards by $10\%$.  This correction is in excellent agreement with the conclusions from our study. 
As in the case of the $\Ysz$--$\Nt$ relation, the covariance between $\Mwl$ and $\Nt$ may be somewhat overestimated.
For instance, \cite{anguloetal12} measured $\Mwl$ within $\avg{\Rf|\Nt}$, whereas $\Mwl$ in R09 comes from
directly fitting a halo model to the shear data.  Consequently, the \citet{anguloetal12} value likely needs to be
reduced for the impact of aperture-induced covariance.\footnote{It is worth emphasizing that whether the correlation coefficient
needs to be reduced by this effect or not has nothing to do with problems in the \citet{anguloetal12} analysis, but rather detailed
differences in how the mass is estimated between \citet{johnstonetal07} and the simulations of \citet{anguloetal12}: the
weak lensing mass measurements are just not identical in detail.  In other cases, such as for the $\Ysz$ measurement that
we discussed earlier, such aperture-induced covariance needs to be explicitly included as per the simulations of
\citet{anguloetal12}.}
However, orientation effects due to optical cluster selection does appear
to induce some covariance that can lead to $\approx 6\%$ overestimates of cluster masses if this affects
is unaccounted for (J\"org Dietrich, private communication), and the \citet{nohcohn12} value for the correlation coefficient
is not affected by aperture-induced covariance.
Overall, it seems clear that a $\approx 10\%$ downwards correction
to the R09 masses due to covariance between $\Mwl$ and $\Nt$
is a plausible explanation for the shift in the mass-scale of maxBCG galaxy clusters needed by
our self-consistent model of multi-variate cluster scaling relations.  We note, however, that such a shift
is also within the systematic error allotted due to cluster miscentering and source photometric redshift errors.

In this context, it is also worth emphasizing that while much of the work addressing the P11-opt data has focused on 
the \citet{johnstonetal07} scaling relation, this is not well justified.  
As was first pointed out by \citet{mandelbaumetal08}, one expects significant photometric
redshift corrections relative to the raw data from \citet{johnstonetal07}.   These correction arise because the lens and source
populations in the SDSS are overlapping, which tends to dilute the weak lensing signal.  Moreover, the inverse critical
surface density varies quickly with source redshift when $z_{source}\approx z_{lens}$, making weak lensing masses
more sensitive to photometric errors.
R09 applied these corrections, and combined the corrected
\citet{sheldonetal09} and \citet{johnstonetal07} analysis with that of \citet{mandelbaumetal08b} to place priors on the 
$M$--$\Nt$ relation of maxBCG galaxy clusters, including this effect.
These corrections {\it should} be applied.\footnote{In fact, there is additional evidence that the original
\citet{johnstonetal07} masses suffer from photometric redshift biases.  
\citet{rozoetal09b}
demonstrated that there is large, systematic-driven evolution in the richness--mass relation of the maxBCG galaxy clusters,
seen both in the X-rays and velocity dispersions \citep{beckeretal07,rykoffetal08a}.
This evolution is not seen in the weak lensing data.  Systematic redshift errors like those pointed out by \citet{mandelbaumetal08b}
have the right magnitude and sense to explain the lack of evolution in the weak lensing data.}
Moreover, there are now several independent stacked weak lensing mass calibrations of the maxBCG
galaxy clusters \citep{mandelbaumetal08b,simetetal12,baueretal12}, all of which are consistent with the R09 scaling relation.
Importantly, all these studies are also subject to the covariance-induced Malmquist bias noted above.


\subsection{Preferred Set of Scaling Relations}

We have demonstrated that the bias-corrected R09 and V09 cluster scaling relations are fully
self-consistent;  not only are the observable--mass relations derived from optical and X-ray selected
cluster catalogs consistent with one another, the optical and X-ray spatial abundance functions are consistent
with each other and with cosmological expectations.  
Because the $\Lx$--$M$ and $\Ysz$--$M$ scaling relations predicted from the bias-corrected
maxBCG and V09 scaling relations are in good agreement with each other, we combine them
to arrive at our preferred estimates for these relations.  These are summarized in Table \ref{tab:final_rel},
along with our preferred set of scaling relations for $M$--$\Nt$, $L_X$--$\Nt$, $\Ysz$--$\Nt$, and $\Ysz$--$L_X$.
For this last set of scaling relations we do not attempt to combine the maxBCG and V09 bias corrected
results: rather, we rely on the measurement that most directly probes each scaling relation.  We emphasize, however,
that this set of scaling relations is still fully self-consistent with the quoted uncertainties.
The pivot point for $L_X$--$M$ and $\Ysz$--$M$ is set to $4.4\times 10^{14}\ \msun$
has been chosen so as to decorrelate the amplitude and slope of the joint scaling relations.
We expect a conservative estimate of the systematic uncertainty due to the hydrostatic
bias and weak lensing bias corrections is $10\%$ in mass, which we expect is best modeled
as a top-hat distribution rather than a Gaussian.  This error corresponds to a 
$\pm 0.16$ error in the amplitude for $\Lx$--$M$, and a $\pm 0.17$ uncertainty in the amplitude for $\Ysz$--$M$.
We now employ our preferred set of scaling relations to predict the cluster masses for each of the CLASH systems,
and to compare the masses recovered from these scaling relations to those from various works in the literature.


\subsection{CLASH Predictions}

Our final set of scaling relations in Table \ref{tab:final_rel}
fits an extra-ordinary amount of data, and satisfies a large variety
of highly non-trivial internal consistency constraints.  As such, we believe it can
provide a critical
low-redshift foundation for the study of cluster scaling relations and their evolution. In addition, these relations provide
a clear target for the CLASH \citep{postmanetal12}  experiment, which seeks to provide very high precision masses
for a small subset of X-ray and lensing selected galaxy clusters. 
In Table \ref{tab:clash}, 
we have provided predictions for the cluster masses for each of the $z\leq 0.4$ CLASH galaxy clusters based
on their X-ray luminosity, as quoted in M10.  We have limited ourselves to $z\leq 0.4$ clusters to minimize
the impact of redshift evolution in our scaling relations, for which we assume self-similar evolution,
$L_X\propto E^2(z)$, as appropriate for soft X-ray band cluster luminosities.
For the few CLASH systems not in M10, Mantz shared with us
X-ray luminosities derived using the same data analysis pipeline as in M10 (Mantz, private communication).
Note that to make these predictions, we utilize the distribution $P(M|\Lx)$
rather than $P(\Lx|M)$.  The two are related as described in Appendix A of paper II, and we have assumed 
a slope of the halo mass function of $\beta=4$, as appropriate for $\approx 10^{15}\ \msun$ galaxy clusters.

Unfortunately, the large scatter in the $\Lx$--$M$ relation implies that our predicted masses are highly uncertain,
as there is an irreducible $\pm 25\%$ intrinsic scatter.
Predictions using the SZ signal are significantly more precise, but there is only one $z\leq 0.4$
CLASH system with SZ measurements in the P11 sample, Abell 2261, which we discuss more fully below.  
$\Ysz$ estimates for these galaxy clusters is very desirable for the purposes of tightening our theoretical
predictions.
We also note that all of our mass predictions
have an overall systematic floor of $\pm 10\%$.
As emphasized in paper II, it is worth remembering that
the mass estimates from scaling relations are only as good as the input data
used to estimate the cluster masses.  
For instance, in the MCXC cluster catalog \citep{piffarettietal11},
clusters A 383, A 209, A 1423, A 611, and MACS J1532,  all have luminosities that differ from the ones quoted
above by $30\%$ or more.  Should these luminosities be correct, then our mass predictions would necessarily
be biased by the corresponding amount.

At this time, there are only two clusters with published masses from a full-lensing analysis of CLASH data:
Abell 2261 and MACS J1205-08.  
Our prediction for Abell 2261 is $M_{500c}=1.07\pm 0.28\ (stat) \times 10^{15}\ \msun$
based on its X-ray luminosity, and $M_{500c}=1.16\pm 0.14\  \times 10^{15}\ \msun$ based
on its SZ signal (both have an additional $\pm 10\%$ top-hat systematic uncertainty).  For the
latter, we again assume self-similar evolution, with $\Ysz \propto E(z)^{2/3}$, and 
we have applied the expected aperture
correction due to the shift in mass calibration between our work and that of P11.
\citet{coeetal12} quote virial masses and concentrations rather than $M_{500c}$, so we use their best fit model to convert
their numbers to $M_{500c}$. They find $M_{500c}=1.34\pm 0.13 \times 10^{15}\ \msun$, in excellent agreement
with either of our predictions.
By comparison, the mass of A2261 in P11--- which P11 obtains based on their $Y_X$ measurements --- is $M_{500c}=0.64\pm 0.04$,
is in tension with the \citet{coeetal12} value at $5\sigma$.  

Turning to MACS J1205-08, our predicted masses from $L_X$ and $\Ysz$ are 
$M_{500c}=1.26 \pm 0.34 \times 10^{15}\ \msun$ and $M_{500c}= 1.36\pm 0.20 \times 10^{15}\ \msun$
respectively.  For the $L_X$-based mass, we relied on the X-ray luminosity from P11.
The reported mass in \citet{umetsuetal12} corresponds to $M_{500c}=1.01 \pm 0.15 \times 10^{15}\ \msun$,
in excellent agreement with both our X-ray and SZ derived cluster mass estimates.
The corresponding mass in P11 is $M_{500c}=1.08 \pm 0.24$, also in good agreement with the CLASH
measurement.  We emphasize that the predicted uncertainties in our SZ masses are $12\%$ and $15\%$
respectively, so the agreement between our predicted masses and those from the CLASH collaboration
is non-trivial.  It should also be noted that \citet{umetsuetal12} find good agreement between their lensing mass
estimates and direct hydrostatic mass estimates from a joint analysis of \chandra\ and \xmm\ data and
an independent analysis of SZ data as per \citet{mroczkowski11}.


\begin{deluxetable}{lccc}
\tablewidth{0pt}
\tablecaption{$M_{500c}$ Predicted Masses for $z\leq 0.4$ CLASH Clusters}
\tablecomment{
X-ray luminosities are all from M10, or from Mantz (private communication).  
The quoted mass uncertainty for our predicted masses ($4^{th}$ column, based
on our preferred $L_X$--$M$ relation) is
statistical only, including the intrinsic scatter in the $M$--$\Lx$ scaling relation, and uncertainty
in the X-ray luminosity estimates.
There is an additional overall $\pm 10\%$ systematic uncertainty in the cluster masses
that scales all clusters uniformly. 
\vspace{0.1in}
}
\tablehead{
Cluster & $z$ & $\Lx\ (10^{44}\ \mbox{ergs/s})$ & $M_{500c}\ (10^{15}\ \msun)$ }
\startdata
A 383 &  0.187 &    5.9 $\pm$    0.2 &   0.69 $\pm$   0.18 \\ 
A 209 &  0.206 &    8.6 $\pm$    0.3 &   0.87 $\pm$   0.23 \\ 
A 1423 &  0.213 &    6.2 $\pm$    0.4 &   0.70 $\pm$   0.19 \\ 
A 2261 &  0.224 &   12.0 $\pm$    0.4 &   1.07 $\pm$   0.28 \\ 
RX J2129 &  0.234 &    9.9 $\pm$    0.5 &   0.94 $\pm$   0.25 \\ 
A 611 &  0.288 &    7.5 $\pm$    0.4 &   0.75 $\pm$   0.20 \\ 
MS 2137 &  0.313 &   12.3 $\pm$    0.4 &   1.02 $\pm$   0.27 \\ 
MACS J1532 &  0.345 &   19.8 $\pm$    0.7 &   1.36 $\pm$   0.36 \\ 
RX J2248 &  0.348 &   30.8 $\pm$    1.6 &   1.80 $\pm$   0.49 \\ 
MACS J1931 &  0.352 &   19.7 $\pm$    1.0 &   1.36 $\pm$   0.36 \\ 
MACS J1115 &  0.352 &   14.5 $\pm$    0.5 &   1.10 $\pm$   0.29 \\ 
MACS J1720 &  0.391 &   10.2 $\pm$    0.4 &   0.86 $\pm$   0.23 \\ 
MACS J0429 &  0.399 &   10.9 $\pm$    0.6 &   0.89 $\pm$   0.24 
\vspace{0.05in}
\enddata
\label{tab:clash}
\end{deluxetable}



\subsection{Predicted Masses for V09, M10, and P11}

\begin{figure}
\begin{center}
\hspace{-0.2in}
\scalebox{1.2}{\plotone{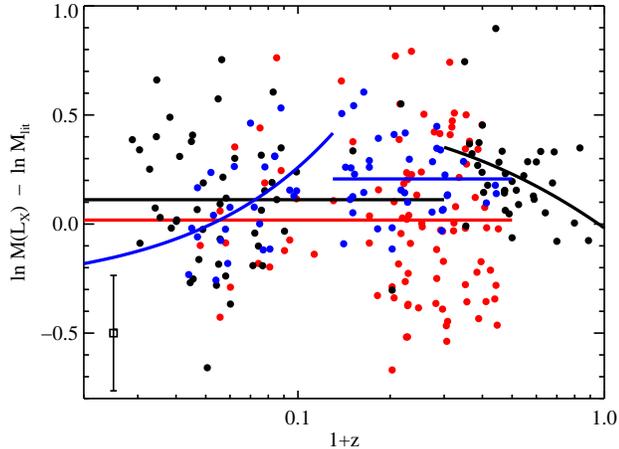}}
\caption{The mass offset $\ln \avg{M|L_X} - \ln M_{lit}$ between the expectation value of the
cluster mass as per our scaling relations, and the mass quoted in V09 (black points), M10 (red points),
and P11 (blue points).   The lines show the mean mass for each of the various cluster sub-samples, including
a low/high redshift partition of the V09 and P11 cluster samples.  The length of each of the lines shows
the redshift range used to define each cluster subsample.  We plot a power-law model in $1+z$ whenever
the best fit power-law has a non-zero slope at more than $2\sigma$ significance.  Otherwise, we plot
the inverse-variance weighted mean.
The square with error bars in the lower left hand
corner is to illustrate the typical uncertainty in the offsets, which is entirely driven by the intrinsic scatter of
the $M$--$L_X$ relation.
}
\label{fig:xray_mass_comparison}
\end{center}
\end{figure}

Just as we predicted the cluster masses for the CLASH cluster sample, we can use our best fit $L_X$--$M$
relation to derive cluster masses for each of the galaxy clusters in V09, M10, and P11.  
Our results are summarized in Figure \ref{fig:xray_mass_comparison}, where we 
plot $\ln \avg{M|L_X} - \ln M_{lit}$ versus cluster redshift for each of the three cluster samples:
V09 (black points),
M10 (red points), and P11 (blue points). 
To evolve the scaling relation away from $z=0.23$, we assume a self-similar evolution $L_X \propto E^2(z)$.
In deriving these masses, we use the X-ray luminosity
quoted in each work.

As expected, our recovered masses are higher than those from V09:
$11\% \pm 4\%$ for the low redshift sample, and $22\%\pm 5\%$ for the high redshift cluster sample.
Note that our putative bias-corrected V09 scaling relation assumed a $15\%$ hydrostatic
bias, which would naturally result in a $\approx 20\%$ offset.  However, upon combining this
scaling relation with maxBCG one, the mass scale decreases slightly, leading to the smaller
low redshift bias estimated above.  
Turning to M10, we find that our predicted masses are in excellent agreement with theirs:
the mass offset is $\Delta \ln M = 2\% \pm 3\%$,  and there is no evidence
of evolution in the mass offset.   M10 notes that a \chandra\ calibration update that appeared as
the work was being published lowers their masses relative to the published values by $\approx 11\%$,
which would result in a net offset $\Delta \ln M = 13\% \pm 3\%$.

Turning to the P11 data set, for the low redshift
sample ($z\leq 0.13$) we see a clear redshift trend in the data, with the mass offset
increasing with increasing redshift.  Whether this trend will remain as deeper data comes out remains to be seen. 
If we ignore this evolution and simply compute the mean mass offset,
we find $\Delta \ln M = 9\% \pm 6\%$.
For the high redshift cluster sample ($z>0.13$), we no longer see systematic evolution in the mass offset, finding
$\Delta \ln M = 21\% \pm 4\%$.    We warn, however, that because these clusters were also SZ-selected, the SZ selection
may pick up the more massive clusters at fixed X-ray luminosity.  Such a selection would tend to increase the X-ray masses
from P11, thereby reducing the apparent offset relative to our mass calibration.  To test for this possibility, we also computed
the mass offset relative to the X-ray masses in \citet{prattetal09}, for which we find $\Delta\ln M = 28\% \pm 0.05\%$.

The mass offset between our predicted masses and those of M10 and P11 are very surprising.  Naively, we would 
have expected the sum of the offsets between our masses and those
in M10 and P11 to be consistent with the M10--P11 offset, but this is not the case.
That is,
despite the fact that at at $z\approx 0.3$ we find that the M10 masses are higher than those in P11 by $\approx 45\% \pm 6\%$,
when we look at the full M10 and P11 cluster samples, our masses are only $\approx 20\%$ higher than those in P11,
and consistent with M10.  The difference is due to the particular cluster sub-sample on which we based the M10--P11
comparison: for this sub-sample, our predicted masses are $18\%\pm 6\%$ {\it lower} than those in M10, and
$29\%\pm 4\%$ higher than those in P11.   Evidently, exactly which clusters are used for these type of comparisons
can dramatically impact the conclusions we draw.   One possible explanation for this difference is that the SZ selection
of the P11 cluster sample implies that high-$L_X$, low $M_{gas}$ systems are under-represented in the P11 sample.
For instance, the P11 sample may be picking out the tail of high $M_{gas}$ clusters in M10 at fixed $L_X$.
Qualitatively, this is exactly analogous
to the problem of the MCXC sub-sample of maxBCG galaxy clusters: starting from a sample selected on a noisy estimator ($L_X$),
the interpretation of sub-samples selected using higher-quality mass proxies can be subject to important selection effects due
to observable covariance.
Quantitatively addressing this question, however, would require extensive simulations with covariant scatter between 
$L_X$, $\Ysz$, and $M_{gas}$, which is beyond the scope of this work.  This does, however, illustrate that the
type of naive comparisons that we are performing in this section and the next can in principle be subject to significant
selection effects due to covariance between cluster observables.


\subsection{Comparison to Other Weak Lensing Masses}

We now perform a similar exercise to the one above, but focusing on masses derived from weak lensing analyses 
instead.
The LoCuSS collaboration\footnote{http://www.sr.bham.ac.uk/locuss/index.php} 
have set out to accurately calibrate cluster scaling relations by providing detailed weak lensing
masses of a sample of $\approx 100$ X-ray selected galaxy clusters.  Their largest compilation of weak lensing masses to
date is that from \citet{okabeetal10}.  In a recent work, the Planck collaboration compared their X-ray
derived masses to those from \citet{okabeetal10}, finding that the latter are $\approx 20\%$ lower than the X-ray masses
derived by the Planck team \citep{planck12_wl}.  This measurement stands in stark contrast to the solution
advocated here, as we have argued that the P11 masses are too low by $\approx 20\%$.

\begin{figure}
\begin{center}
\hspace{-0.2in}
\scalebox{1.2}{\plotone{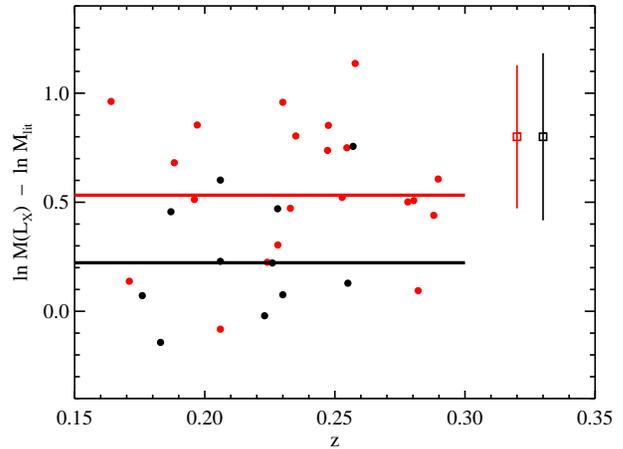}}
\caption{The mass offset $\ln \avg{M|L_X} - \ln M_{lit}$ between the expectation value of the
cluster mass as per our scaling relations, and the weak lensing mass quoted in \citet[][red points]{okabeetal10} 
and \citet[][black points]{mahdavietal08}.  The lines show the inverse-variance weighted mean for each sample.
The squares with error bars in the upper right hand
corner illustrate the typical uncertainties in the offsets, which can be dominated by the weak lensing
uncertainties.
}
\label{fig:wl_mass_comparison}
\end{center}
\end{figure}

In Figure \ref{fig:wl_mass_comparison}, we compare our predicted masses from $L_X$ to those quoted 
in \citet[][red points]{okabeetal10}.
There is a very large offset between our predictions and their masses, $\Delta \ln M = 0.53 \pm 0.07$,  nearly a factor
of two.  This value is somewhat higher than the expected offset of $\approx 0.4$ derived from the naive
sum of the P11 offset relative to our predictions, and the offset quoted in \citet{planck12_wl}.

As was demonstrated in paper II, the mass calibration used by P11 results in
strong ($> 4\sigma$) tension between the observed X-ray luminosity function, and the cosmological expectations for a WMAP7
cosmology.  Further lowering 
the cluster masses can only increase this tension.   Should the \citet{okabeetal10} 
masses be correct, the X-ray luminosity function in \citet{bohringeretal04} would need to be corrected upwards by a factor of 3 to 4, 
implying the REFLEX catalog is $\approx 70\%$ incomplete.
We find this level of incompleteness much too high to be plausible, so we expect there remains an unknown source
of systematic bias in the weak lensing measurements of \citet{okabeetal10}.  
Interestingly, \citet{okabeetal10} measured the CLASH cluster A 2261 as part of their Subaru weak lensing
campaign.  For this cluster, they find
$M_{500c}=8.14\pm 1.17\times 10^{14}\ \msun$.\footnote{We symmetrized their error bar.}
This mass is indeed biased low relative to the CLASH value, with the net mass offset being 
$\Delta \ln M = 0.50 \pm 0.18$.  


We also compare our overall mass calibration to that of \citet{mahdavietal08}, relying on luminosity estimates from M10
to perform the comparison.  This cluster sample is identical to that
of \citet{hoekstra07}, but updated with improved photometric redshifts for source galaxies.   
Our results are shown in Figure \ref{fig:wl_mass_comparison} using black points.  The mean mass offset
is $\Delta \ln M = 0.22\pm 0.12$, with our predicted masses being larger than those of \citet{hoekstra07}.
Given the large uncertainties in the weak lensing masses, however, the two data sets are statistically consistent
with each other.  

For completeness, we have also searched for clusters in common to \citet{mahdavietal08} and \citet{okabeetal10}, finding
6 such systems.  One cluster, Abell 209, might be an outlier, but it is difficult to tell with such a small sample size.  
The mean mass offsets with and without Abell 209
are $-0.27 \pm 0.29$ and $-0.37 \pm 0.18$ respectively.  Note that despite the large offsets between the two works, the
scatter is such that based on these 6 systems we cannot conclude that the two data sets are inconsistent
with each other.

\citet{hoekstraetal11} have extended the cluster sample from \citet{mahdavietal08} to include low mass, high redshift
systems.  We assume a concentration parameter $c_{500}=R_{500c}/R_{2500c}=3$ to convert their reported $M_{2500c}$ masses
to $M_{500c}$, but note that our conclusions are sensitive to the assumed value of $c_{500}$.  The value we
adopt here is consistent with X-ray observations \citep[e.g.][]{vikhlininetal06} and numerical simulations 
\citep[e.g.][]{bhattacharyaetal11}.  A reasonable estimate for the systematic uncertainty in the mass corrections is $\pm 10\%$.
With our fiducial correction, we find that the mean mass offset between the two works is $\Delta \ln M = 13\%\pm 13\%$.
Thus, our predicted masses are fully consistent with those in \citet{hoekstraetal11}. We caution, however, that
there is very large scatter in this comparison, reflecting the large errors in the weak lensing masses: 12\% of the systems
in that work have a reported mass that is within $1\sigma$ of zero, and a full $68\%$ of the systems are within $2\sigma$
of zero.  

The final work we consider here is that of \citet{umetsuetal11}, who combined magnification and shear measurements
to derive constraints on five galaxy clusters: A 1689, A 1703, A 370, CL0024+17, and RXJ1347-11.  The inverse variance
weighted mean offset relative to our predictions (using luminosities in M10 or provided by Adam Mantz, private communication)
is $\Delta \ln M = -0.42 \pm 0.14$, with our predicted masses being lower.  
This offset reflects the fact that there are two systems with fairly large offsets: A370, with $\Delta \ln M = -0.85 \pm 0.29$,
and CL0024+17, with $\Delta \ln M = -1.26\pm 0.30$.   The offset for CL0024+17 ($L_X=2.3\pm 0.1$)
is so large that it is the only system we have encountered where our predicted mass was more than $3\sigma$
away from the reported mass.  Both A370 and CL0024+17 are truly exceptional systems.
A370 was one of the first cluster lenses ever identified \citep{soucailetal87,soucailetal87b}, and may be the most massive 
lensing cluster in the Universe \citep{broadhurstetal08,umetsuetal11}.  It also exhibits bimodality in its mass distribution,
both in X-rays and lensing \citep{richardetal10}, with the two components separated in redshift 
by $\approx 3,000\ \mbox{km/s}$ \citep{defilippisetal05}.  CL0024+17 is also known as an exceptional 
lensing system \citep{zitrinetal09}, and appears to be a recent cluster merger
along the line of sight, leading to high weak lensing mass and a highly diffuse gas distribution \citep{umetsuetal10}.
Given that these two clusters are known to be some of the most effective lenses in the Universe, 
and the small sample size of the cluster sample, we are loathe to derive any conclusions from this comparison.

A final large compilation of homogeneously analyzed weak lensing masses of galaxy clusters is that of \citet{ogurietal12}.
Unfortunately, these systems are typically not very X-ray bright: they were all first identified in the optical.  Consequently,
X-ray luminosities for these systems are not easily available in the literature, preventing us from performing a comparison
with that work.  A summary of the mass offsets for both X-ray and weak lensing samples is presented in Table \ref{tab:moffset}.
As noted in the previous section, however, we emphasize that a more meaningful comparison would require a careful
treatment of selection effects to avoid reaching biased conclusions because of intrinsic observable
covariance.


\begin{deluxetable}{lll}
\tablewidth{0pt}
\tablecaption{Mass Offset Between our Predicted Masses and Values from the Literature}
\tablecomment{
Mean mass offset between our predicted masses using $L_X$, and those reported in
the literature.  All means are inverse-variance weighted. 
}
\tablehead{
Work & $\ln \avg{M|L_X} - \ln M_{lit}$ & No. of Clusters \\
& & in Sample }
\startdata
V09 ($z\leq 0.3$) & $0.11\pm 0.04$ &  49      \\
V09 ($z> 0.3$) & $0.22\pm 0.05$ & 36       \\
M10 & $0.02\pm 0.03$ & 95       \\
P11 ($z\leq 0.13$) & $0.09\pm 0.05$ & 24       \\
P11 ($z> 0.13$) & $0.21\pm 0.04$ & 38       \\
\citet{okabeetal10} & $0.53\pm 0.07$ & 21       \\
\citet{mahdavietal08} & $0.22\pm 0.12$ & 11       \\
\citet{hoekstraetal11} & $0.13\pm 0.13$ &  25      \\ 
\citet{umetsuetal11} & $-0.42\pm 0.14$ & 5
\vspace{0.05in}
\enddata
\label{tab:moffset}
\end{deluxetable}



\subsection{The Thermal SZ Power Spectrum}

As final check on our preferred set of scaling relations, 
we consider the implications of our $\Ysz$--$M$ relation on the thermal SZ (tSZ) spectrum.
Specifically, we compare the predicted amplitude of the tSZ power spectrum derived
from our preferred scaling relation to observations from the South Pole 
Telescope \citep[SPT,][]{reichardtetal11}.  
Following their convention, we characterize the
amplitude of the thermal power spectrum  
in terms of its value relative to the fiducial model of
\citet{shawetal10}.  That is, by definition, the \citet{shawetal10} model corresponds to $\Atsz=1$.  
The value of the best fit amplitude $\Atsz$ to the SPT data depends 
on whether or not one allows for possible covariance with
the Cosmic Infra-red Background (CIB).  The best
fit values are $\Atsz=0.70\pm 0.21$ when assuming no covariance, 
and $\Atsz=0.60\pm 0.24$ when marginalizing
over any possible covariance.  

These values are to be compared with those predicted
from our best fit $\Ysz$--$M$ relation assuming self-similar evolution.  \citet{shawetal10} found
that the tSZ amplitude for the
\citet{arnaudetal10} $\Ysz$--$M$ model is $A_{tSZ,Arnaud}=1.27$, in moderate tension
with the data.   
The principal difference between our joint best fit $\Ysz$--$M$ relation and that of \citet{arnaudetal10}
is the normalization, which differ in amplitude by $\Delta a_{sz|m} = 0.465$.
Since $\Atsz\propto \Ysz^2$, the predicted amplitude $\Atsz$ from our joint
$\Ysz$--$M$ scaling relation is $\Atsz=0.50\pm 0.10$, in excellent agreement with the data.
The offsets relative to the data with and without CIB covariance are 
$0.4\sigma$ and $0.9\sigma$ respectively.


\section{Summary and Discussion}
\label{sec:discussion}

We have demonstrated that the bias-corrected R09 and V09 cluster scaling relations are fully
self-consistent;  not only are the observable--mass relations derived from optical and X-ray selected
cluster catalogs consistent with one another, the optical and X-ray spatial abundance functions are consistent
with each other and with cosmological expectations.  

Table \ref{tab:final_rel} summarizes our preferred set of scaling relations.  
As usual, the scaling relations with mass are dominated by systematic uncertainties, which we estimate at the $\pm 10\%$
level on the cluster mass.    Note that other than the joint V09+maxBCG $\Lx$--$M$ and
$\Ysz$--$M$ scaling relations, we have not attempted to combine all of our data sets: we have simply selected
the cluster sample that probes each cluster scaling relation most directly.  
In principle, one could use the
model from Appendix A in paper II to perform a full likelihood analysis on the full collection of data sets.
Such an analysis would recover not only constraints on the individual scaling relations, but also on the
various correlation coefficients.   Such an analysis, however, is beyond
the scope of this work. In fact, we think of our work as a necessary precursor to such an analysis: given
the surprising results presented in P11-opt concerning the $\Ysz$--$\Nt$ relation, it was necessary to
investigate whether there even existed a model that could give a good fit to all of the data currently available.
Otherwise, it would
make little sense to attempt to combine these various cluster samples.

Concerning the $\Ysz$--$\Nt$ relation, 
there are several differences between our analysis and that of P11-opt.  First, all of our scaling relations
are self-consistently propagated using the probabilistic model described in Appendix A of paper II 
\citep[see also Appenidx C in][]{whiteetal10}, or through
explicit Monte Carlo methods.  In addition, P11-opt
used the \citet{arnaudetal10} model for the $\Ysz$--$M$ scaling relation when predicting the $\Ysz$--$\Nt$
relation, whereas we explicitly constrained the $\Ysz$--$M$ relation for each of the three X-ray data sets
we considered, and used this as the basis of our analysis.
It is worth noting that because the $\Ysz$/$Y_X$ ratio in
P11 at $z=0.23$ is higher than that derived in \citet{arnaudetal10}, the tension between our P11(z=0.23) prediction
and the P11-opt data is even stronger than that in the original P11-opt paper.

The tension between our own prediction for the P11(z=0.23) $\Ysz$--$\Nt$ relation and the P11-opt data has three
important contributions, all involving mass.  First, the P11 masses appear to be biased low relative to those from
other X-ray works, particularly V09 and M10.  In addition, we expect the masses in these two works should be further
corrected due to hydrostatic bias, at the $\approx 5\%-15\%$ level (see below).  
Finally, we had to lower the maxBCG weak lensing masses by $10\%$, a correction
that is at least partly sourced by intrinsic covariance between weak lensing mass and cluster richness at
fixed mass.  We note, however, that this systematic shift is also within the expected systematic
error due to photometric redshift uncertainties and cluster miscentering.
Shifting these values around by up to $\pm 10\%$ is acceptable, but larger shifts start to introduce tension in various places: 
either the observable--mass relations from optical and X-ray catalogs will fail to be consistent with one another, or the spatial 
abundance of galaxy clusters become inconsistent with cosmological expectations.  

Throughout this work, we adopted a 15\% hydrostatic bias (at fixed aperture, 21\% total bias) 
when modeling the V09 bias-corrected scaling relations.
This value is well supported by simulation results spanning more than twenty 
years \citep[e.g.][]{evrard90, evrard96, cen97, rasiaetal06, nagaietal07a,lauetal09,battagliaetal11,rasiaetal12}.
Curiously, when we directly compare the cluster masses derived from our preferred scaling relation to those
quoted in V09, the net offset at low and high redshifts was only $\approx 10\%$ and $20\%$ respectively,
suggesting an overall total hydrostatic bias close to $\approx 10\%$ at fixed aperture, also well within
the range found in numerical simulations.

If hydrostatic bias were in fact significantly higher, say $\approx 30\%$ or higher as seen in some simulations,
then the agreement between the V09 and maxBCG data sets would be a 
fortuitous coincidence; such a large hydrostatic bias would have to be cancelled by some other unknown bias
in the V09 data that tends to increase cluster masses.   In such a scenario, the P11(z=0.23) data would in 
fact be consistent or close to consistent with maxBCG observations.   Note, however, that the puzzle of the high $\Ysz/Y_X$
ratio from the P11(z=0.23) data noted in paper I would remain.  Since this latter scenario requires a conspiracy of errors,
and still leaves an open question unanswered, we much prefer our proposed solution.
As for the weak lensing optical masses, the bias in the R09 scaling relation may be explained as a combination of
Malmquist bias due to covariance between weak lensing masses and $\Nt$.

In summary, we believe we have been able to present a solution to the puzzle posed in P11-opt: that is, we have constructed
a set of cluster scaling relations that satisfies all the internal requirements for self-consistency.  These
conditions are numerous and non-trivial.  In fact, we have seen these conditions tightly constrain deviations from our proposed
solution: neither the mass scale of the maxBCG or V09 galaxy clusters can be altered by much more than $\approx 10\%$
without introducing tension in either some cluster scaling relation, or with cosmological expectations for the cluster spatial abundance.
We note too that our recovered $\Ysz$--$M$ scaling relation is fully consistent with the amplitude of the thermal SZ effects
as measured in the SPT data \citep{reichardtetal11}.

Having derived our preferred $L_X$--$M$ and $\Ysz$--$M$ relations from these arguments, we have used them 
to estimate cluster masses for several cluster samples in the literature,
and to predict cluster masses for each of the $z\leq 0.4$ CLASH systems.  Despite the large statistical uncertainty in our
predicted cluster masses --- which are dominated by the intrinsic scatter in the $M$--$L_X$ relation --- we can average over
many clusters to test the overall level of systematic mass offset between that of our favored $L_X$--$M$ relation, and
the masses reported in the literature.  These mass offsets are summarized in Table \ref{tab:moffset}.  
We caution, however, that these mass offsets can be subject to important selection effects
due to intrinsic covariance between cluster observables.  Importantly, we were able to also derive masses from the SZ
observations in P11 for the two galaxy clusters that have been published so far by the CLASH collaboration, 
Abell 2261 \citep{coeetal12} and MACS J1205-08 \citep{umetsuetal12}.  In both cases, we find excellent agreement between
our predicted masses and the CLASH results.  We emphasize that this agreement if highly non-trivial: both our predicted masses
and the results from the CLASH collaboration quote statistical uncertainties of order $\approx 10\%-15\%$.

At this time,  the largest difference between the V09 and maxBCG catalogs
involves the amplitude of the SZ signal, see for instance Figures \ref{fig:yszn_sys}, \ref{fig:yszm} and \ref{fig:yszlx}: 
in all cases,
boosting the P11-opt $\Ysz$ measurement by $\approx 20\%$ --- particularly for low mass objects --- 
would result in better agreement with the V09 data 
set and the P11-X measurements.  We emphasize, however, that such a boost is not necessary at this point; 
the current data set is self-consistent in a statistical sense. Interestingly, a $20\%$ boost of the $\Ysz$--$\Nt$ amplitude
corresponds to less than a $2\sigma$ (statistical only) shift in the $\Ysz$--$\Nt$ relation recovered from the P11-opt
data.  With deeper \planck\ data, the statistical error in the $\Ysz$--$\Nt$ relation will shrink significantly; should the amplitude
not shift upwards as the statistical precision of the measurements improves, this new data could very well rule out our
proposed solution for $\sim 10^{14}\ \msun$ galaxy clusters.

\acknowledgements The authors would like to thank the organizers of the Monsters Inc. workshop at KITP, 
supported in part by the National Science Foundation under Grant No. PHY05-51164, where
this collaboration was started.  The authors also gratefully acknowledge T. Biesiadzinski for sharing their
systematic corrections to $\Ysz$ in the maxBCG data, and A. Mantz for sharing his X-ray luminosities 
for those systems not published in M10.
ER gratefully acknowledges the hospitality of the AstroParticle and Cosmology 
laboratory (APC) at the Universit\'e Paris Diderot, where part of this work took place. 
ER is funded by NASA through the Einstein Fellowship Program, grant PF9-00068. 
AEE acknowledges support from NSF AST-0708150 and NASA NNX07AN58G. 
JGB gratefully acknowledges support from the Institut Universitaire de France.
A portion of the research described in this paper was carried out at the Jet Propulsion Laboratory, 
California Institute of Technology, under a contract with the National Aeronautics and Space Administration.
This work was supported in part by the U.S. Department of Energy contract to SLAC no. DE-AC02-76SF00515.

\newcommand\AAA[3]{{A\& A} {\bf #1}, #2 (#3)}
\newcommand\PhysRep[3]{{Physics Reports} {\bf #1}, #2 (#3)}
\newcommand\ApJ[3]{ {ApJ} {\bf #1}, #2 (#3) }
\newcommand\PhysRevD[3]{ {Phys. Rev. D} {\bf #1}, #2 (#3) }
\newcommand\PhysRevLet[3]{ {Physics Review Letters} {\bf #1}, #2 (#3) }
\newcommand\MNRAS[3]{{MNRAS} {\bf #1}, #2 (#3)}
\newcommand\PhysLet[3]{{Physics Letters} {\bf B#1}, #2 (#3)}
\newcommand\AJ[3]{ {AJ} {\bf #1}, #2 (#3) }
\newcommand\aph{astro-ph/}
\newcommand\AREVAA[3]{{Ann. Rev. A.\& A.} {\bf #1}, #2 (#3)}

\bibliographystyle{apj}
\bibliography{mybib}



\end{document}